\documentclass[twocolumn]{aastex62}
\usepackage{bm}
\usepackage{amsmath}
\submitjournal{ApJ}

\shorttitle{Alfv\'en wave-driven wind from evolved stars}
\shortauthors{Yasuda et al.}
\begin{document}

\title{Alfv\'en wave-driven wind from RGB and AGB stars}

\correspondingauthor{Yuki Yasuda}
\email{yuki.yasuda@sci.hokudai.ac.jp}

\author[0000-0003-1093-0052]{Yuki Yasuda}
\affiliation{Division of Earth and Planetary Sciences, Faculty of Science,\\
Hokkaido University, Sapporo 060-0810, Japan}

\author{Takeru K. Suzuki}
\affiliation{School of Arts and Sciences, University of Tokyo, 
3-8-1, Komaba, Meguro, Tokyo, 153-8902, Japan}

\author{Takashi Kozasa}
\affiliation{Division of Earth and Planetary Sciences, Faculty of Science,\\
Hokkaido University, Sapporo 060-0810, Japan}
\affiliation{Department of Cosmosciences, Graduate School of Science, 
Hokkaido University, Sapporo 060-0810, Japan}



\begin{abstract}
We develop a magnetohydrodynamical model of Alfv\'en wave-driven wind 
in open magnetic flux tubes piercing the stellar surface of 
Red Giant Branch (RGB) and Asymptotic Giant Branch (AGB) stars, 
and investigate the physical properties of the winds. The model 
simulations are carried out along the evolutionary tracks of stars 
with initial mass in the range of 1.5 to 3.0 
$M_{\odot}$ and initial metallicity $Z_{\rm ini}$=0.02. 
The surface magnetic field strength 
being set to be 1G, we find that the 
wind during the evolution of star can be classified into 
the following four types; the first is the wind with 
the velocity higher than 80 km s$^{-1}$ in the RGB 
and early AGB (E-AGB) phases; the second is the wind 
with outflow velocity less than 10 km s$^{-1}$ seen around 
the tip of RGB or in the E-AGB phase; the 
third is the unstable wind in the E-AGB and thermally 
pulsing AGB (TP-AGB) phases; the fourth is the stable 
massive and slow wind with the mass-loss rate higher than 
10$^{-7} M_{\odot}$ yr$^{-1}$ and the outflow velocity lower than 20 
km s$^{-1}$ in the TP-AGB phase. The mass-loss rates in 
the first and second types of wind are two or 
three orders of magnitude lower than the values evaluated by 
an empirical formula. The presence of massive and slow wind 
of the fourth type suggests the possibility that the 
massive outflow observed in TP-AGB stars could be attributed to 
the Alfv\'en wave-driven wind.

%(253 words) --> km s^-1 wo 1 word , Mun yr ^-1 wo 1 word to surunara 249 words

\end{abstract}



%%%%%%%%%%%%%%%%%%%%%%%%%%%%%%%%
\section{Introduction}
In the post main-sequence phase, the envelope of low and intermediate star expands during 
the contraction of central core; in the Red Giant Branch 
(RGB) and the early Asymptotic Giant Branch (E-AGB) phases, 
the stellar radius reaches typically several tens of the solar radius,  
and several hundreds of the solar 
radius in the thermally pulsing AGB (TP-AGB) phase. 
As stars ascend along the RGB and the AGB, the mass-loss rate becomes larger, which has been 
revealed by various observations \citep[e.g.,][]{mes09,sch13,ram14}, and the 
mass-loss rate in these phases increases with decreasing the surface gravity \citep{jud91}. 
The mass loss influences the observable properties such 
as the spread of the horizontal branch (HB) in the Hertzsprung-Russell (HR) diagram 
of globular clusters \citep[e.g.,][]{cat00} and the 
number ratio of TP-AGB to RGB stars (N$_{\rm TP-AGB}$/N$_{\rm RGB}$) on 
the color magnitude diagram \citep{ros14,ros16}.

Furthermore, Third Dredge-Up (TDU) and Hot Bottom Burning (HBB) experiencing on TP-AGB, 
whose efficiencies depend on the mass-loss rate, affect the surface abundances of C, N, 
and O elements, the effective temperature, and the surface gas density 
\citep[e.g.,][]{mar02,ven10,tas17}. Consequently, the mass-loss rate influences the chemical 
composition and amount of dust formed around TP-AGB stars 
\citep[e.g.,][]{ven12,ven18} as well as 
the spectral appearance of the stars. In the stellar evolution calculations, the 
empirical mass-loss formulae of \citet{rei75} and \citet{sch05}  have been often 
employed on the RGB and the AGB \citep[e.g.,][]{wei09,cri09,cri11,mar13}. 
On the other hand, recently \citet{ros14,ros16} have claimed that 
an enhanced mass loss in comparison with the rates evaluated by the 
empirical formulae is favorable in the pre-dust phase on the TP-AGB to reproduce 
N$_{\rm TP-AGB}$/N$_{\rm RGB}$ in nearby galaxies with metallicity ranging from 
[Fe/H] = -1.59 to -0.56. 
 
So far, as the promising physical processes related to mass loss on the RGB or/and AGB, the following three 
processes have been considered; the dissipation of Alfv\'en 
wave which has been believed to act in both phases as long as the chromospheric 
structure is present; levitation of the atmosphere induced by stellar pulsation; radiation pressure 
acting on dust grains and the resulting momentum transfer from dust to gas through 
collisions. The second one alone cannot induce the outflow observed in AGB stars 
\citep{sch03}. However the hydrodynamical models of pulsation-enhanced dust-driven 
wind that combines the latter two 
%ones 
can induce massive outflow from AGB stars. 
In the early models, the third one was treated  
as a simple function of gas temperature \citep{woo79,bow88}. 
Then the models have been further developed 
by considering the formation process of dust concretely 
for carbon-rich (C--rich) AGB stars \citep[e.g.,][]{fle92,hof97,yas12}.
Although recent high-resolution observation shows that the pulsation-enhanced 
dust-driven wind mechanism actually works in evolved AGB stars, 
the applicability for oxygen-rich (O--rich) AGB stars 
has yet to be explored \citep[e.g.,][]{ohn16}.
Thus, it has been desirable to investigate how the first 
process affects the mass-loss event on the RGB and the AGB using 
magnetohydrodynamical (MHD) models \citep[e.g.,][]{vid06}. 

The mass loss in the RGB phase has been considered to be induced by an 
outward-directed flux of Alfv\'en waves in the stellar atmosphere
\citep[e.g.,][]{bel75,hai80,har80}. As mentioned by \citet{hol83}, 
in the steady--state models, the massive slow winds can be reproduced by 
adding the energy not only in the subsonic region but also in the 
supersonic region; the energetics of the Alfv\'en wave-driven wind needs 
to be treated as accurately as possible. Furthermore it is unclear whether the stellar winds  
continuously stream out in the framework of the steady--state models. 
Then the dynamical MHD simulations for the RGB phase were performed 
by \citet{suz07} applying the wind model originally developed for the Sun 
that reproduces the physical properties of the observed solar corona and wind 
\citep{suz05,suz06}. \citet{suz07} showed that the model for the RGB stars can 
generate the massive wind 
whose mass-loss rate $\dot{M}$ varies in time from 10$^{-10}$ to 5 $\times$ 10$^{-7} M_{\odot}$ yr$^{-1}$ 
However, as pointed out by \citet{air10}, the wind model does not include 
magnetic diffusion terms in the induction equation. If the magnetic diffusion is taken 
into account, the Alfv\'en waves dissipate more rapidly near the stellar surface, 
which reduces the energy supply in the outer region. As a result, the wind could be expected 
to be less dense in realistic situations. One of the main purposes in this 
paper is to extend the model of \citet{suz07} by including effects of the magnetic 
diffusion. 

In addition, the evaluation of the radiation field is crucial for the energetics of the 
stellar wind. In particular, the radiative cooling and heating processes  
in the region with the temperature lower than ten thousands directly affects  
the formation of chromosphere in the Alfv\'en wave-driven wind. 
However, in the above mentioned dynamical models for the Alfv\'en wave-driven 
wind \citep{suz07,air10}, the radiative heating process has not been treated. 

In order to investigate Alfv\'en wave-driven winds in  
open magnetic flux tubes piercing the stellar surface of  RGB and AGB stars, 
we extend the MHD model of \citet{suz07} as follows; first 
we include the magnetic diffusion terms via Joule dissipation and ambipolar diffusion. 
Second, evaluating the radiation field based on a modified Unno--Kondo 
method by \citet{win97}, we take into account radiative heating and cooling processes.  
Then, simulating the stellar winds for model stars whose stellar parameters are 
taken from the stellar evolution calculations, we examine the dynamical properties of 
winds along the evolution tracks.

This paper is organized as follows. In Section 2, we describe our wind model, 
focusing on the parts extended from the model of \citet{suz07}. Then, after the wind 
types are classified in Section 3.1, Section 3.2 presents the results of the simulated 
winds on the RGB and the AGB of stars with initial mass $M_{\rm ini}$ = 1.5, 2.0, and 
3.0 $M_{\odot}$ and initial metallicity $Z_{\rm ini}$ = 0.02 as a representative of 
solar metallicity. In Section 3.3, the transition of wind type is investigated as 
a function of stellar parameters. In Section 4.1, we discuss the results of simulations, 
analyzing the change of wind properties along the evolution tracks of stars. 
Then we compare the results of the numerical simulations with those based on observations 
for RGB and E-AGB stars (Section 4.2.1) and for TP-AGB stars (Section 4.2.2). The summary is 
presented in Section 5.

\section{Model}

\subsection{Magnetohydrodynamical model} 
In order to construct the Alfv\'en wave-driven wind model for evolved 
stars with extended atmospheres,  we extend the one-dimensional (1D) MHD 
simulation code developed for the RGB stars in \citet{suz07}. Here we describe the 
extensions of the code.

We set an open magnetic flux tube by a filling factor $f$, and  
the functional form, following \citet{kop76} and \citet{suz13}, is given by 
\begin{eqnarray}
f(r)=\frac{e^{(r-r_{{\rm 0}}-h_{{\rm 1}})/h_{{\rm 1}}}+f_{{\rm 0}}-(1-f_{{\rm 0}})/e}
{e^{(r-r_{{\rm 0}}-h_{{\rm 1}})/h_{{\rm 1}}}+1}, 
\label{filling_f}
\end{eqnarray}
where $r$ is the radial distance from the center, $f_{{\rm 0}}$ is a filling 
factor at the stellar surface of the radius $r_{{\rm 0}}$, and $h_{{\rm 1}}$ 
is a typical height of closed magnetic loops \citep[e.g., see Figure 1 of][]{suz13}. 
In the simulations, $f_{{\rm 0}}$ is set to be 1/1000 as a standard value 
\citep[e.g.,][]{suz13}, and we adopt $h_{{\rm 1}}$ = 0.6 $h_{{\rm p}}$  with 
the pressure scale height $h_{{\rm p}}$ at the stellar surface. 
Then the radial component of the magnetic field $B_{r}$ is given as 
\begin{eqnarray}
B_{r}(r)=B_{r,{\rm 0}}\frac{f_{{\rm 0}}r_{{\rm 0}}^{2}}{f(r)r^{2}}
\label{dif_Br}
\end{eqnarray}
by the conservation of magnetic flux, and the radial magnetic field at the stellar surface 
$B_{r,{\rm 0}}$ is set to be 1000G as a standard value so that the averaged field strength 
contributed from the open field regions $f_{{\rm 0}}B_{r,{\rm 0}}$ = 1G.

Using the filling factor $f$, we solve the equations for the conservation laws of mass, 
momentum, and energy, and the induction equation that are expressed as 
%\begin{flalign}
\begin{eqnarray}
& &\frac{d\rho}{dt}+\frac{\rho}{r^{2}f}\frac{\partial}{\partial r}\left(r^{2}fv_{r}\right)=0,\label{gas_cont_eq}\\
& &\rho\frac{dv_{r}}{dt}=-\frac{\partial p}{\partial r}
-\frac{1}{8\pi r^{2}f}\frac{\partial}{\partial r}\left(r^{2}fB_{\perp}^{2}\right)
+\frac{\rho v_{\perp}^{2}}{2r^{2}f}\frac{\partial}{\partial r}\left(r^{2}f\right)
\nonumber\\
& &\;\;\;\;\;\;\;\;\;\;\;\;\;\;\;\;-\rho\frac{GM_{\ast}}{r^{2}},
\label{gas_mom_eq_radial}\\
& &\rho\frac{d}{dt}\left(r\sqrt{f}v_{\perp,i}\right)=\frac{B_{r}}{4\pi}
\frac{\partial}{\partial r}\left(r\sqrt{f}B_{\perp,i}\right),
\label{gas_mom_eq_perp}\\
& &\rho\frac{d}{dt}\left(\epsilon+\frac{v^{2}}{2}+\frac{B^{2}}{8\pi\rho}-\frac{GM_{\ast}}{r}\right)
+\frac{1}{r^{2}f}\frac{\partial}{\partial r}
\left\{r^{2}f\left[\left(p\right.\right.\right.\nonumber\\
& &\left.\left.\left.
+\frac{B^{2}}{8\pi}\right)v_{r}-\frac{B_{r}}{4\pi}\bm{B}\cdot\bm{v}\right]\right\}
-\frac{1}{4\pi r^{2}f}\frac{\partial}{\partial r}\left[r\sqrt{f}\eta\sum_{i=1}^{2}
B_{\perp,i}\right.\nonumber\\
& &\left.\frac{\partial}{\partial r}\left(r\sqrt{f}B_{\perp,i}\right)\right]
+\frac{1}{r^{2}f}\frac{\partial}{\partial r}\left(r^{2}fF_{c}\right)
+q_{R}=0,\label{gas_ene_eq}\\
& &\frac{\partial B_{\perp,i}}{\partial t}=\frac{1}{r\sqrt{f}}\frac{\partial}{\partial r}\left[
r\sqrt{f}(v_{\perp,i}B_{r}-v_{r}B_{\perp,i})\right. \nonumber\\
& &\;\;\;\;\;\;\;\;\;\;\;\;\;\;\;\;\left.+\eta\frac{\partial}{\partial r}\left(r\sqrt{f}B_{\perp,i}\right)\right],
\label{induc_eq}
\end{eqnarray}
%\end{flalign}
where symbols are almost the same as those used in \citet{suz07}; $t$, $\rho$, $\bm{v}$, $p$, and 
$\bm{B}$ are time, mass density, velocity, thermal pressure, and magnetic field strength, 
respectively, and subscripts $r$ and  $\perp,i$ denote radial and $i$th tangential component, 
respectively; $d/dt$ and $\partial/\partial t$ denote Lagrangian and Eulerian derivatives, respectively; 
$\epsilon$ is the specific internal energy and is given by $\epsilon=p/[(\gamma-1)\rho]$ for an ideal gas 
with the specific heat ratio $\gamma$; $G$ is the gravitational constant; $F_{c}=\kappa_{{\rm 0}}T^{5/2}(dT/dr)$ is 
the thermal conductive flux by Coulomb collisions, 
which is effective only in high temperature regions with temperature $T$ $\geq$ 10$^{4}$ K, 
where $\kappa_{{\rm 0}}=10^{-6}$ g cm s$^{-3}$ K$^{-7/2}$; $q_{R}$ represents the net radiative 
cooling rate described in Section 2.2; the magnetic diffusivity $\eta$ is the 
sum of the Joule diffusion $\eta_{{\rm J}}$ and the ambipolar diffusion  $\eta_{{\rm AD}}$, 
each of which is given by  
\begin{eqnarray}
\eta_{{\rm J}}\!&=&\!\frac{c^{2}m_{e}\nu_{en}}{4\pi e^{2}n_{e}}\!\simeq 2.3 \!\!\times\!\! 10^{2}\frac{1}{x_{e}}
\left(\frac{T}{{\rm K}}\right)^{1/2}\text{cm$^{2}$s$^{-1}$}\;\;\;\;\;\;\;\;
\label{def_eta_J}\\
\text{and}& &\nonumber\\ 
\eta_{{\rm AD}}\!&=&\!\frac{B^{2}}{4\pi\chi\rho_{i}\rho_{n}}\!\simeq \!2.1 \!\!\times\!\! 10^{-16}
\frac{(B/\text{G})^{2}}{\left[\rho/(\text{g \!\!\! cm$^{-3}$})\right]^{\!2}\!\!\!x_{e}}
\;\text{cm$^{2}$s$^{-1}$}\!\!\!,\;\;\label{def_eta_AD}
\end{eqnarray} 
%\begin{eqnarray}
%\eta_{{\rm J}}\!&=&\!\frac{c^{2}m_{e}\nu_{en}}{4\pi e^{2}n_{e}}\!\simeq 2.3 \!\!\times\!\! 10^{2}\frac{1}{x_{e}}
%\left(\frac{T}{{\rm K}}\right)^{1/2}\text{cm$^{2}$s$^{-1}$}\;\;\;\;\;\;\;\;
%\label{def_eta_J}\\
%\text{and}& &\nonumber\\ 
%\eta_{{\rm AD}}\!&=&\!\frac{\!(\rho_{n}/\!\rho)^{2}\!B^{2}\!}{\!4\pi\chi\rho_{i}\rho_{n}\!}\!\simeq \!2.1 \!\!\times\!\! 10^{-16}
%\!\frac{(B/\text{G})^{2}}{\left[\rho/(\text{g \!\!\! cm$^{-3}$})\right]^{\!2}\!\!\!x_{e}\!}
%\;\text{cm$^{2}$s$^{-1}$}\!\!\!,\;\;\label{def_eta_AD}
%\end{eqnarray}
respectively, where $c$ is the the speed of light, $e$ is the elementary charge, 
and subscripts $e$, $i$, and $n$ denote electron\footnote{Although $e$ is used for the elementary charge and 
electron, we keep this expression because it does not cause confusion.}, ion, and neutral species, 
respectively; $m_{e}$ is the mass of electron; $\nu_{en}$ is the collision frequency between electron 
and neutral species and is given as $\nu_{en}=n_{n}<\sigma_{en}v_{en}>$ with 
the number density of neutral species  $n_{n}$, 
the electron-ion collision cross section $\sigma_{en}$, and the relative velocity between ion and neutral 
species $v_{en}$, and where the angle bracket means average over the velocity space. $\chi$ is given as 
$\chi=<\sigma_{in}v_{in}>/(m_{i}+m_{n})$, where $\sigma_{in}$, $v_{in}$, and $m_{i}$ ($m_{n}$) are 
the ion-neutral collision cross section, the relative velocity between ion and neutral species, and 
the mass of ion (neutral) species, respectively; $x_{e}=n_{e}/n_{\rm H}$ is the ionization degree 
that is defined as the ratio of $n_{e}$ to  
%the total number density $n$. 
the number density of hydrogen nuclei $n_{\rm H}$; 
We evaluate the numerical values of right-handed sides of equations~(\ref{def_eta_J}) 
and (\ref{def_eta_AD}) by setting $<\sigma_{en}v_{en}>$ $=$ 8.3 $\times$ 
10$^{-10}$ $T$/K cm$^{3}$ s$^{-1}$ and $<\sigma_{in}v_{in}>$ $=$ 
1.9 $\times$ 10$^{-9}$ cm$^{3}$ s$^{-1}$, referring to equations 14 and 12, respectively, of \citet{dra83}. 
Equation (9) is valid if and only if $x_{e}$ $\ll$ 1. As $x_{e}$ approaches to unity, $\rho_{n}$ 
becomes lower than $\rho$, and equation~(\ref{def_eta_AD}) overestimates $\eta_{\rm AD}$. 
Furthermore, in the circumstellar envelope of the model star on the RGB and on the 
E-AGB, the ion-neutral collision time typically exceeds the maximum period of the perturbation waves 
(set at the stellar surface) in the temperature region of T $>$ 20000 K. 
Thus we turn off $\eta_{\rm AD}$ for $x_{e}$ $>$ 0.8. 
Here we note that the Hall effect does not work in this 1D simulations 
without dependence on the two tangential components. To estimate $x_{e}$, we disregard the 
second and higher ionized metal species for simplicity. Then $x_{e}$ is approximately expressed as 
\begin{eqnarray}
x_{e}&=&\frac{n_{e}}{n_{\rm H}}\nonumber\\ 
&=&\frac{n_{p}}{n_{\rm H}}+\frac{n_{\rm He^{+}}}{n_{\rm H}} +2\frac{n_{\rm He^{++}}}{n_{\rm H}}
+\sum_{j=1}^{J}A_{j}\left(\frac{R_{c1}^{j}}{R_{1c}^{j}}+1\right)^{-1}\!\!\!\!, \;\;\;\;\;\;
\label{def_x_e}
\end{eqnarray}
where  $n_{p}$, $n_{\rm He^{+}}$, and $n_{\rm He^{++}}$ are the number densities 
of hydrogen ions, first ionized helium ions, second ionized helium ions, respectively; 
$A_{j}$ is the abundance of $j$th metal species relative to hydrogen;  
$R_{1c}^{j}$ and $R_{c1}^{j}$ are the photoionization rate and the radiative 
recombination rate, respectively,  between the ground and continuum states for the $j$th metal species. 
We derive the ratios $n_{p}/n_{\rm H}$,  
$n_{\rm He^{+}}/n_{\rm H}$, and $n_{\rm He^{++}}/n_{\rm H}$
referring to the scheme of \citet{har84} and Harper (2014, priv.\ comm.). 
%\textbf{(Harper (2014, priv.(J\(B comm.))}. 
The ratio  $R_{1c}^{j}/R_{c1}^{j}$ is set as 
\begin{eqnarray}
\frac{R_{1c}^{j}}{R_{c1}^{j}}=\frac{1}{n_{e}T}\left(\frac{2\pi m_{e}k_{\rm B}T}{h^{2}}\right)^{3/2}
\left[WT_{\rm eff}e^{-h\nu_{\i,0}/(k_{\rm B}T_{\rm eff})}\right.
\nonumber\\
+\left.W_{\rm gal}T_{\rm gal}e^{-h\nu_{\i,0}/(k_{\rm B}T_{\rm gal})}\right],\;\;\;\;\;\;\;\;\;\;\;\;\;\;\;\;\;\;\;\;
\;\;\;\;
\label{def_R_ratio}
\end{eqnarray}
where $k_{\rm B}$ and  $\nu_{\i,0}$ are the Boltzmann constant and the frequency of the photoionization edge.  
Here we assume that the radiation field of star is approximated by optically thin blackbody 
radiation characterized by the effective temperature $T_{\rm eff}$ and a geometric dilution factor $W$  
defined as 
\begin{eqnarray} 
W=\frac{1}{2}\left[1-\sqrt{1-\left(\frac{R_{\ast}}{r}\right)^{2}}\right]
\end{eqnarray} 
where $R_{\ast}$ is the stellar radius. In equation~(\ref{def_R_ratio}), we also include a galactic 
ionization term due to the interstellar radiation field (ISRF) particularly in the ultraviolet (UV) region.  
Referring to the work of \citet{mat83}, we set the temperature of the UV component $T_{\rm gal}$ as 7500 K  
and the dilution factor of the component $W_{\rm gal}$ as 10$^{-14}$.

\subsection{Radiative cooling and heating} 
The net radiative cooling term $q_{R}$ appearing in equation~(\ref{gas_ene_eq}) is expressed 
by the difference of the radiative cooling rate $q_{R,{\rm cool}}$ and the radiative heating rate 
$q_{R,{\rm heat}}$ as follows; 
\begin{eqnarray}
q_{R}=q_{R,{\rm cool}}-q_{R,{\rm heat}}.
\label{def_q_R}
\end{eqnarray}

In this paper we take the following simple treatment for the cooling term $q_{R}$. We introduce the 
following three temperature regimes; $T$ $<$ $T_{{\rm low}}$, $T_{{\rm low}}$ $\leq$ $T$ $<$ $T_{{\rm high}}$, 
and $T_{{\rm high}}$  $\leq$ $T$, where $T_{{\rm low}}$ = 6000 K and $T_{{\rm high}}$ = 10000 K. In the high 
temperature regime with $T$ $\geq$ $T_{{\rm high}}$, we basically adopt optically thin cooling rate given by   
$q_{R,{\rm cool}} = n_{p}n_{e}\Lambda$, where $\Lambda$ is the cooling 
function for the optically thin plasma with solar metallicity \citep{sut93}
\footnote{On the TP-AGB where the surface abundances of C, N, and O considerably 
change due to TDU and HBB, this estimate of the cooling rate based on (scaled) 
solar abundances is problematic in the hot component of 
chromosphere with $T$ higher than 15000 K. However the chromospheric structure is not formed 
on this evolutionary phase as shown in Section 3. Therefore we expect that 
the disregard of the dependency of the cooling rate on the chemical components 
in the high temperature regime does not affect the simulation results.}. 
However, this treatment sometimes overestimates the cooling in optically-thick region. In order to 
avoid this overestimation, we adopt a "cap" taken from the empirical radiative 
rate = $4.5$ $\times$ 10$^{9}$ $\rho$ erg cm$^{-3}$ s$^{-1}$ of the 
solar chromosphere \citep{and89,mor04} for $q_{R,{\rm cool}}$; 
\begin{eqnarray}
q_{R,{\rm cool},{\rm high}}=\min(n_{p}n_{e}\Lambda, 
4.5\! \times 10^{9}\! \rho\;\;{\text{erg cm\!$^{-2}$ s\!$^{-1}$}}).\;\;\;\;\;
\label{def_q_R_cool_high}
\end{eqnarray}
In the low-temperature range, $T$ $<$ $T_{{\rm low}}$,  
the radiative cooling rate $q_{R,{\rm cool}}$ is the same as in the constant opacity model for 
pulsation-enhanced dust-driven wind from TP-AGB stars \citep[e.g.,][]{tas17}. The value in this case 
$q_{R,{\rm cool},{\rm low}}$ is given by using the Rosseland mean gas opacity at the stellar surface 
$\kappa_{{\rm surf}}$ derived by the stellar evolution calculations, and is simply expressed as  
\begin{eqnarray}
q_{R,{\rm cool},{\rm low}}=4\pi \kappa_{{\rm surf}} B(T)=4\kappa_{{\rm surf}}\sigma T^{4}, 
\label{def_q_R_cool_low}
\end{eqnarray}
where $B(T)$ and $\sigma$ are the frequency integrated Planck function and the Stefan--Boltzmann constant, 
respectively. In the intermediate range, $T_{{\rm low}}$ $\leq$ $T$ $<$ $T_{{\rm high}}$, we 
interpolate between $q_{R,{\rm cool},{\rm low}}$ and $q_{R,{\rm cool},{\rm high}}$ in a logarithmic manner:
\begin{eqnarray}
q_{R,{\rm cool},{\rm int}}\!=\!\left[q_{R,{\rm cool},{\rm low}}(T_{{\rm low}})\right]^{1-\alpha}
\!\left[q_{R,{\rm cool},{\rm high}}(T_{{\rm high}})\right]^{\alpha}\!\!\!\!,\;\;\;\;\;\;
\label{def_q_R_cool_int}
\end{eqnarray}
where $\alpha=\log(T/T_{{\rm low}})/\log(T_{{\rm high}}/T_{{\rm low}})$. Although equation~(\ref{def_q_R_cool_int}) 
is a quite simplified prescription, we suppose that it can give a reasonable estimate because the timescale of 
the radiative cooling is almost logarithmically monotonic in the temperature range of 6000 K $<$ $T$  
$<$ 10000 K \citep[Figure 11 of][]{woi96}.

The radiative heating rate $q_{R,{\rm heat}}$ is simply estimated by using the frequency integrated mean intensity $J$
as follows;
\begin{eqnarray}
q_{R,{\rm heat}}=4\pi\kappa_{{\rm surf}}J.
\label{def_q_R_cool}
\end{eqnarray}
This equation is a reasonable approximation in optically thick region unless the scattering is dominant.
It should be cautioned that using the Rosseland mean opacity $\kappa_{\rm surf}$ could underestimate the 
heating rate in optically thin region, which will be investigated in our future work. 
We estimate the frequency integrated mean intensity $J$ in equation~(\ref{def_q_R_cool}) by using 
the Unno-Kondo method in the gray approximation \citep{win97}: 
\begin{eqnarray}
J&=&\frac{L_{\ast}}{8\pi^{2}}\left[\frac{1}{(1+\mu_{r}(R_{{\rm out}}))R_{{\rm out}}^{2}}
+\frac{\mu_{r}(R_{{\rm out}})}{R_{{\rm out}}^{2}}-\frac{\mu_{r}}{r^{2}}\right]\nonumber\\
& &+\frac{L_{\ast}}{8\pi^{2}}\left[\frac{3}{2}\int_{r}^{R_{{\rm out}}}\frac{dr'}{r'^{2}}\left(\rho\kappa_{{\rm surf}}
+\frac{2\mu_{r}(r')}{r'}\right)\right]\;\;\;\; \label{UK}
\end{eqnarray}
where $R_{{\rm out}}$ is the radial position of the outer boundary, and $\mu_{r}$ is the cosine of the separation 
angle, which is derived by integrating 
\begin{eqnarray}
\frac{\partial \mu_{r}^{3}}{\partial r}=\frac{\rho\kappa_{{\rm surf}}(1-5\mu_{r}^{2})}{4}
+\frac{3\mu_{r}(1-\mu_{r}^{2})}{r}\label{sep_angle}
\end{eqnarray}
from the inner boundary $R_{{\rm in}}$ with the boundary condition 
\begin{eqnarray}
\mu_{r}(R_{{\rm in}})=0.\label{inner_BC}
\end{eqnarray}
The above framework to estimate radiation field has been adopted in our previous wind models from TP-AGB stars 
\citep[e.g.,][]{tas17} with the addition of the dust opacity.

\subsection{Input parameters} 
The input parameters of our MHD simulations are the followings: the stellar mass $M_{\ast}$, 
the stellar luminosity $L_{\ast}$, the effective temperature $T_{{\rm eff}}$ as well as  
the gas opacity $\kappa_{{\rm surf}}$, the average magnetic field $f_{{\rm 0}}B_{r,{\rm 0}}$, 
and the fluctuation amplitude $\langle\delta v_{{\rm 0}}\rangle$ at the stellar surface. 

Among the input parameters, the stellar parameters $M_{\ast}$, $L_{\ast}$, $T_{{\rm eff}}$, and $\kappa_{{\rm surf}}$ 
along the evolutionary track for a model star with a given set of initial mass $M_{\rm ini}$ 
and initial metallicity $Z_{\rm ini}$ are calculated by the Modules for Experiments in Stellar 
Astrophysics (MESA) code (version r6208) \citep{pax11,pax13,pax15}. The method of calculation is 
the same as in \citet{tas17}, thus we briefly outline the method here; The standard mixing 
length theory \citep{cox68} is applied and  convection is approximated as a diffusive process 
within the convective region defined by Schwarzshild 
criterion, and the mixing length parameter $\alpha_{{\rm MLT}}$ is set to be 2.0.  The overshooting 
parameter $f$ is set to be 0.014 at all convective boundaries, except for 
the bottom of the He--shell flash region at which  $f$ is set to be 0.008 throughout the 
evolution after the first thermal pulse (TP) \citep{pax11}. In this paper we adopt the CNO-enhanced 
low temperature opacity constructed by using \AE SOPUS tool \citep{mar09}, and apply 
the mass-loss formula of \citet{sch05} in the post main-sequence phase. 

Along the evolutionary track of a model star, the stellar parameters are sampled with the stellar luminosity 
spacing of 10$^{0.05}$--10$^{0.2}$ $L_{\ast}$ on the RGB and on the E-AGB, and with the stellar mass spacing 
$\delta M_{\ast}$ $\simeq$ 0.05 $M_{\odot}$ on the TP-AGB, excluding the periods of thermal pulses. 

Following \citet{suz07}, we set $f_{{\rm 0}}B_{r,{\rm 0}}$ = 1G; the value is not 
so deviated from that estimated for S-type Mira star $\chi$ Cyg by \citet{leb14} (2-3G) and 
those estimated for three C--rich AGB stars by \citet{dut17} (1.1--9.5G). We note that, among 
the three AGB stars, the average value of  the surface magnetic field  strength for IRC+10216 is 
3.8 G while those of RW LMi and RY Dra are lower than 4.4 and 4.8G, respectively. 
%The dependence of $f_{{\rm 0}}B_{r,{\rm 0}}$ on the simulation results is investigated in Section 4.2.

The other parameter, $\langle\delta v_{{\rm 0}}\rangle$, is determined as follows. First we refer to 
\citet{ren77} for the dependence of the excited acoustic flux $F_{a,{\rm 0}}$ on the surface gravity 
$g$ and $T_{{\rm eff}}$: 
\begin{eqnarray}
F_{a,{\rm 0}} \simeq \rho_{{\rm 0}}\langle\delta v_{{\rm 0}}^{2}\rangle c_{{\rm S}}
\propto g^{-0.7}T_{{\rm eff}}^{12},\label{F_depend}
\end{eqnarray}
where $\rho_{{\rm 0}}$ and $c_{{\rm S}}$ are the mass density and the sound velocity at the stellar surface, 
respectively. Then, we get the scaling relation of the surface fluctuation,  
\begin{eqnarray}
\langle\delta v_{{\rm 0},{\rm scale}}\rangle\!=\!\left(\!\frac{\rho_{{\rm 0}}}{\rho_{{\rm 0},\odot}}\!\right)^{\!\!-0.5}
\!\!\left(\!\frac{g}{g_{\odot}}\!\right)^{\!-0.35}\!\!\left(\!\frac{T_{{\rm eff}}}{T_{{\rm eff},\odot}}\!\right)^{\!\!5.75}
\!\!\!\!\langle\delta v_{{\rm 0},\odot}\rangle.\;\;\;\;\label{def_del_scale}
\end{eqnarray}
Here the symbol $\odot$ represents the Sun, $\rho_{{\rm 0},\odot}$ $=$ 2.07 $\times$ 10$^{-8}$ g cm$^{-3}$, 
\footnote{From the web site; http://solarscience.msfc.nasa.gov/} and 
$T_{{\rm eff},\odot}$ $=$ 5.77 $\times$ 10$^{3}$ K 
%, and $\rho_{{\rm 0}}$ is the mass density 
at the stellar surface where the diluted optical depth is 2/3 in the initial static atmosphere of our MHD model. 
The value of $\langle\delta v_{{\rm 0},\odot}\rangle$ is set to 1.0 km s$^{-1}$ \citep{suz07}. 
Note that $\langle\delta v_{{\rm 0},{\rm scale}}\rangle$ for the stars on the AGB 
with $R_{\ast}$ $>$ 60 $R_{\odot}$ could exceed the sound velocity $c_{{\rm S}}$ except for the case 
of $T_{\rm eff}$ $\leq$ 3000 K. 
Supersonic fluctuation at the surface simply results in the increase of the radiative 
loss at low altitudes though the structures of the upper atmosphere and envelope are not strongly 
affected \citep{suz13}. Therefore, we adopt 
\begin{eqnarray}
\langle\delta v_{{\rm 0}}\rangle=
{\rm min}\left(\langle\delta v_{{\rm 0},{\rm scale}}\rangle, c_{{\rm S}}\right).\label{def_del_v0}
\end{eqnarray}
The wave spectrum of the fluctuation at the stellar atmosphere is set so as to depend on the inverse of 
frequency according to the recipe described in Section 2.2.3 of \citet{suz07}. Both the longitudinal and 
transverse fluctuations are included, and those amplitudes are set to be 
$\langle\delta v_{{\rm 0}}\rangle$. 

\section{Results}
We present in this section the results of the MHD simulations of the Alfv\'en wave-driven wind from the 
RGB and AGB stars whose stellar parameters derived from the stellar evolution calculations for the 
model stars of $M_{\rm ini}$ = 1.5, 2.0, and 3.0 $M_{\odot}$ with $Z_{\rm ini}$ = 0.02, focusing 
on the wind properties; note that we do not simulate the stellar winds in the HB phase 
(gray colored regions in Figures~\ref{fig1},~\ref{fig7}, and ~\ref{fig8}) as well as those of stars 
descending from the tip of RGB (in particular the region between 
$M_{\ast}$ =2.9983 and 2.9964 $M_{\odot}$ in Figure ~\ref{fig8}). 

The MHD simulations are carried out as follows; Given a set of the input parameters (see Section 2.3), 
the simulation is continued until a continuous gas--outflow is observed at the outer boundary located at 
25  $R_{\ast}$ and the continuous gas--outflow is referred as the stable wind hereafter for convenience; 
otherwise we stop the simulation when the simulation time reaches  80 $R_{\ast}$/$c_{{\rm S}}$ and 
we call the wind observed in this case as the unstable wind hereafter for simplicity.  The mass-loss 
rate $\dot{M}$ and the radial gas velocity $v_{\rm gas}$ that characterize the stable wind are 
calculated by averaging the values at the outer boundary over the latter half of the simulation time. 

In the following subsections, first we summarize the types of winds observed in the simulations 
(Section 3.1), then the properties of wind observed in the simulations are described for each model 
star (Section 3.2). The dependence of the transition of wind properties during the evolution of star 
on the effective temperature  $T_{\rm eff}$ as well as the surface gravity $g$ is clarified 
in Section 3.3. 

\subsection{Classification of wind types} 

The properties of Alfv\'en wave-driven wind change during the evolution of star. Roughly speaking, 
the wind properties are classified into the following four types (Table~\ref{tbl-1:classification}). 
The first type is characterized by warm\footnote{In this paper, we call the warm (cool) wind as the 
wind whose gas temperature is higher than 10$^{5}$ K (lower than $T_{{\rm eff}}$).} 
and fast wind with the radial outflow velocity 
(radial component of the gas velocity) $v_{\rm gas}$  $>$ 80 km s$^{-1}$. 
The wind of the second type exhibits almost time-steady, slow and cool wind with the mass-loss rate 
$\dot{M}$ $<$ 10$^{-10} M_{\odot}$ yr$^{-1}$. 
In the third type, steady wind does not stream out but intermittent winds are driven in a sporadic manner. 
The fourth type is characterized by cool and massive wind with 
$\dot{M}$ $>$ 10$^{-7} M_{\odot}$ yr$^{-1}$. The first and second types are found in the RGB 
and E-AGB phases. The third type is seen in the tip of RGB and the AGB phases while the fourth 
type appears in the late stage of the TP-AGB phase. 
%All the four types of wind do not necessarily appear during the evolution of star. 
The detail is provided for $M_{\rm ini}$ = 1.5, 2.0, and 3.0 $M_{\odot}$ model stars in the following 
subsections. 

\begin{table*}
\centering
\caption{Classification of wind types}
\label{tbl-1:classification}
\begin{tabular}{llll}
\hline
  type name & evolution stage & characteristics & $\log g$ \\
\hline
  first type & RGB and E-AGB & $\dot{M}$ $\leq$ 10$^{-11}$ $M_{\odot}$ yr$^{-1}$ and $v_{\rm gas}$ $>$ 80 km s$^{-1}$
& $>$ 1.1\\
  second type & RGB and E-AGB & 
5 $\times$ 10$^{-11}$ $M_{\odot}$ yr$^{-1}$ $\leq$ $\dot{M}$ $\leq$ 2 $\times$ 10$^{-10}$ $M_{\odot}$ yr$^{-1}$ and  $v_{\rm gas}$ $<$ 10 km s$^{-1}$
& 0.91 to 1.1\\
  third type & tip of RGB and AGB & unstable (or sporadic)  & -1.0 to 1.0 \\
  fourth type & TP-AGB & $\dot{M}$ $>$ 10$^{-7}$ $M_{\odot}$ yr$^{-1}$ and 2.4 km s$^{-1}$ $<$ $v_{\rm gas}$ $<$ 15 km s$^{-1}$
& $<$ -0.4\\
\hline
\end{tabular}
\end{table*}

\subsection{Wind properties of the model stars} 

\subsubsection{$M_{\rm ini}$ = 1.5 $M_{\odot}$ model} 
\begin{figure}
\vspace{0.7cm}
\includegraphics[scale=0.465,angle=-90,trim=10mm 0mm 0mm 0mm]{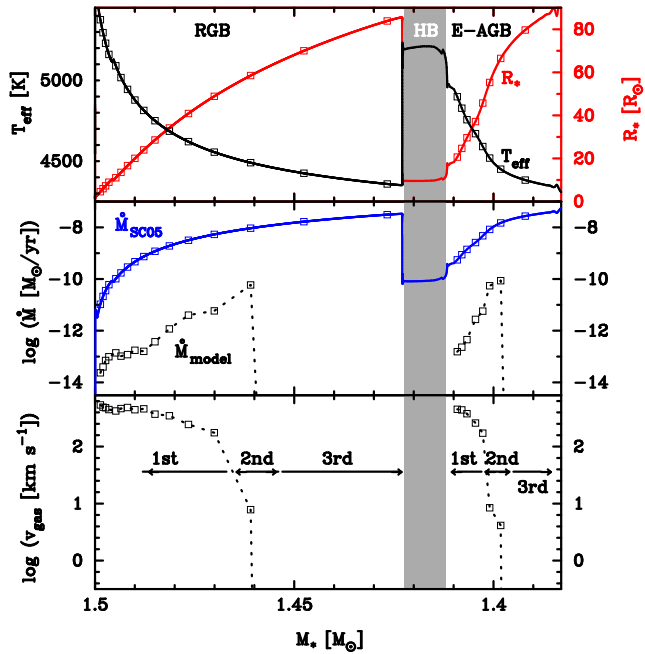}
\vspace{0.7cm}
\caption{Time evolution of stellar parameters and wind properties at the outer boundary from the RGB to 
the E-AGB phase for the model star of $M_{\rm ini}$ = 1.5 $M_{\odot}$. Top panel: the effective temperature 
$T_{\rm eff}$ (black) and the stellar radius $R_{\ast}$ (red). Middle panel : the mass-loss rates of 
our model (dotted black) and \citet{sch05} (blue). Bottom panel: the outflow velocity $v_{\rm gas}$.
(A color version of this figure is available in the online journal.)\label{fig1}}
\end{figure}

Figure~\ref{fig1} shows the time evolution of the effective temperature $T_{\rm eff}$, the stellar 
radius $R_{\ast}$ in units of $R_{\odot}$, the mass-loss rate $\dot{M}$ in units 
of $M_{\odot}$ yr$^{-1}$, and the radial outflow velocity $v_{\rm gas}$ in units of km s$^{-1}$ 
from the RGB to the E-AGB phase; note that the value of $\dot{M}$ and $v_{\rm gas}$ are time-averaged 
values at the outer boundary. The blue line in the middle panel shows the mass-loss rate 
used in the stellar evolution calculation. 

%Immediately after entering the RGB phase, $R_{\ast}$ increases from 4.55 to 7.28 $R_{\odot}$, 
%$M_{\ast}$ decreases from 1.4987 to 1.4973 $M_{\odot}$, $\dot{M}$ increases 
%from 1.46 $\times$ 10$^{-14}$ to 8.62 $\times$ 10$^{-14}$ $M_{\odot}$ yr$^{-1}$, and $v_{\rm gas}$ 
%somewhat decreases from 117 to 88.7 km s$^{-1}$. The change of the wind properties with increasing the stellar radius 
%$R_{\ast}$ partly reflects the increase of the fluctuation amplitude  $\langle\delta v_{{\rm 0}}\rangle$  
%from 2.28 to 3.08 km s$^{-1}$. 
%As the star evolves, both $\dot{M}$ and $v_{\rm gas}$ increase until $R_{\ast}$ becomes 34.2  $R_{\odot}$.

Immediately after entering the RGB phase, 
$R_{\ast}$ increases from 4.55 to 11.0 $R_{\odot}$, $\dot{M}$ increases
from 2.34 $\times$ 10$^{-14}$ to 1.38 $\times$ 10$^{-13}$ $M_{\odot}$ yr$^{-1}$, and 
$v_{\rm gas}$ somewhat decreases from 539 to 421 km s$^{-1}$. 
Then, as the star evolves, both $\dot{M}$ and $v_{\rm gas}$ are not so changed until $R_{\ast}$ becomes 24.0  $R_{\odot}$.

After then, the wind properties drastically change twice on the RGB-phase;   
first at $M_{\ast}$ = 1.4609 $M_{\odot}$ where $R_{\ast}$ = 58.49 $R_{\odot}$ and second 
$M_{\ast}$ = 1.448 $M_{\odot}$ where $R_{\ast}$ = 70.00 $R_{\odot}$.
In particular, this is clearly seen in the temporal variation of the outflow velocity shown in the bottom panel.
Just before the first change, the winds are tenuous and fast; $\dot{M}$  is lower than 
1 $\times$ 10$^{-11} M_{\odot}$ yr$^{-1}$ and $v_{{\rm gas}}$ is higher than 
80 km s$^{-1}$, and the wind is classified as the first type (Section 3.1).  
After the first transition, the chromospheric structures are not always seen, and the wind becomes more 
massive but quite slow; $v_{{\rm gas}}$ is lower than 10 km s$^{-1}$. 
This massive wind is classified as the second type (Section 3.1).  
Then, after the second transition, 
the winds become further slower, and steady wind does not stream out. We call this unstable wind as the third type. 
These two transitions between three wind types are also found in E-AGB phase as shown in the bottom panel in 
Figure~\ref{fig1} ($M_{\ast}$ = 1.401 and 1.392 $M_{\odot}$).
In the following, we show the radial structures of the three types of the winds.

\begin{figure}
\vspace{0.7cm}
\includegraphics[scale=0.42,angle=-90,trim=10mm 0mm 0mm 0mm]{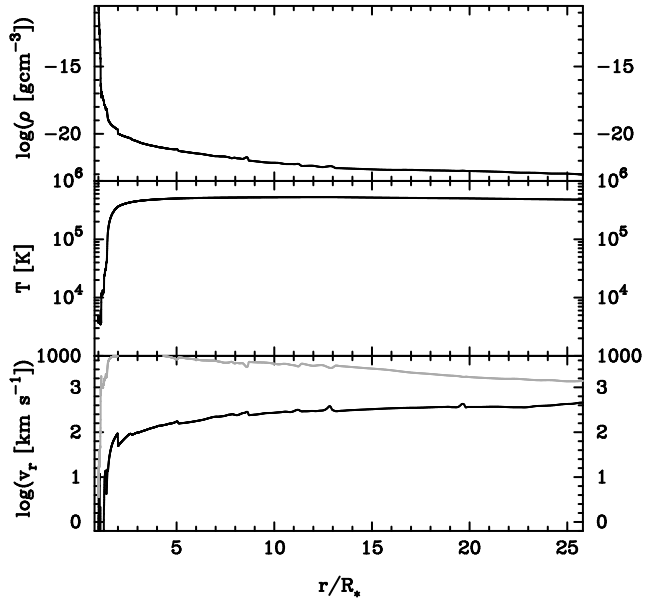}
\vspace{0.7cm}
\caption{Snap shot of the radial structure at $M_{\ast}$ = 1.490 $M_{\odot}$  in the model star of $M_{\rm ini}$ = 1.5 
$M_{\odot}$ at simulation time $t$ = 0.763 yr, which is a typical wind of the first type. 
Top panel: the mass density $\rho$ (solid line) and that of the initial static atmosphere (dashed line). 
Middle panel: the temperature $T$. Bottom panel: the radial 
component of the gas velocity $v_{r}$ (black) and Alfv\'en velocity $v_{\rm A}$ (gray).\label{fig2}}
\end{figure}

Figure~\ref{fig2} depicts the radial structure of a typical example of the first type wind. 
%We show the radial structure of a typical wind at the first type in Figure~\ref{fig2}. 
We can see from the middle panel that the first wind is a warm wind with the gas temperature  
$T$ exceeding 10$^{5}$ K in the region outside 1.46 $R_{\ast}$. The radial gas velocity 
$v_{r}$ exceeds 200 km s$^{-1}$ at 6.56 $R_{\ast}$, but is still sub-Alfv\'enic, as can be seen from the 
bottom panel. As shown in the top panel, $\rho$ is quite low, and does not exceed 10$^{-20}$ g cm$^{-3}$ in 
the region outside 20 $R_{\ast}$. Roughly speaking, the first type wind for the star with $R_{\ast} <$ 41 
$R_{\odot}$ is almost in steady state.

\begin{figure}
\vspace{0.7cm}
\includegraphics[scale=0.42,angle=-90,trim=10mm 0mm 0mm 0mm]{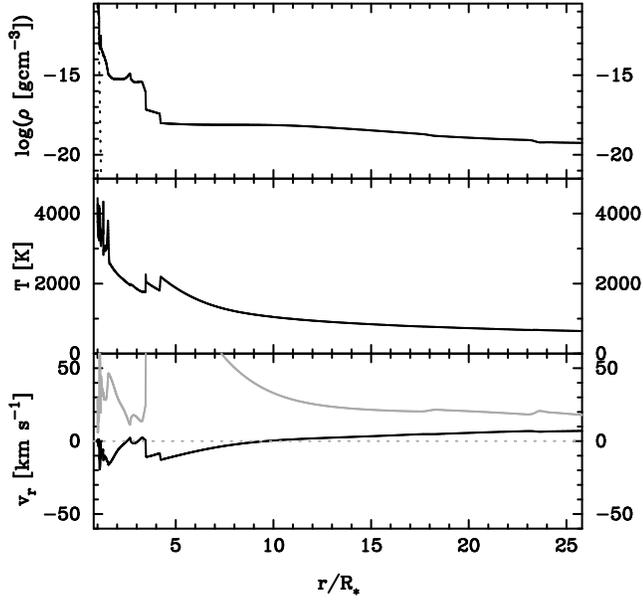}
\vspace{0.7cm}
\caption{Same as Figure 2, but  at $M_{\ast}$ = 1.4609 $M_{\odot}$  in the model star of  $M_{\rm ini}$ = 1.5 
$M_{\odot}$ at simulation time $t$ = 23.2 yr, which is a typical wind of the second type,  
with the addition of zero velocity (dashed line) in bottom panel.\label{fig3}}
\end{figure}

A typical structure of the wind in the second type is shown in Figure~\ref{fig3}. As shown in the middle panel, 
the wind is cool and $T$ is lower than $T_{\rm eff}$.
The wind in the region of $r$ $>$ 9.61 $R_{\ast}$ moves outward as shown in the bottom panel. The 
second type wind is not quasi-steady, and  it is not clear whether the wind could be sustainable 
over longer simulation time or not. The stability of the wind of this type will be discussed in  Section 4.1.

\begin{figure}
\vspace{0.7cm}
\includegraphics[scale=0.42,angle=-90,trim=10mm 0mm 0mm 0mm]{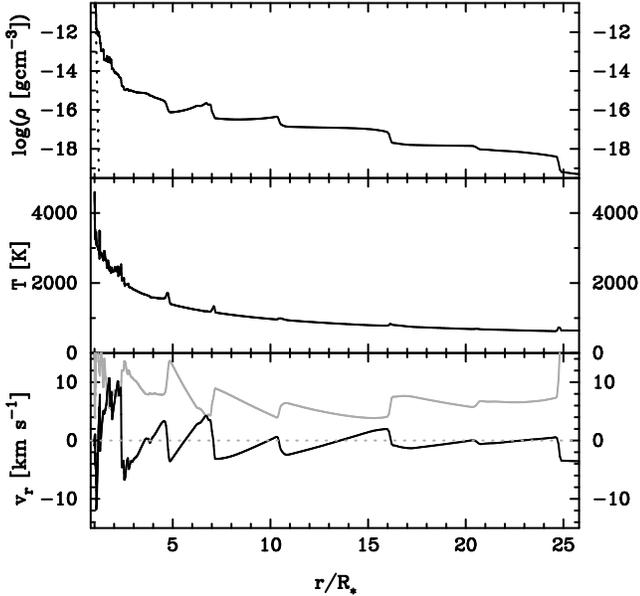}
\vspace{0.7cm}
\caption{Same as Figure 3, but at $M_{\ast}$ = 1.427 $M_{\odot}$ and at simulation time $t$ = 27.0 yr, 
which is a typical wind of the third type.\label{fig4}}
\end{figure}

A structure of the third type wind is presented in Figure~\ref{fig4}. The middle panel 
shows that the chromospheric structure disappears.  The outer atmosphere is levitated due to the injection of the 
Alfv\'en wave, which can be seen in the top panel. However, as shown in the bottom panel, inflows occur in 
intershock regions. 
%an intershock region of $7-11 R_{\ast}$, and the outer region of $r>15 R_{\ast}$. 
Therefore this type of wind is considered not to 
cause any significant mass loss.  The behavior of flowing gas in the E-AGB phase of $M_{\ast} < 1.392 M_{\odot}$ 
is very similar to that of Figure~\ref{fig4}. Thus, the mass loss during the evolution of $M_{\rm ini}$ = 1.5 $M_{\odot}$ 
star seems to stop twice; in the RGB phase of 1.461 $M_{\odot}$ $>$ $M_{\ast}$ $>$ 1.427 $M_{\odot}$ 
%$1.447 M_{\odot} > M_{\ast} > 1.423 M_{\odot}$ 
and in the E-AGB phase of $M_{\ast} < 1.392 M_{\odot}$.  
%$1.392 M_{\odot} > M_{\ast} > 1.382 M_{\odot}$.  

\begin{figure}
\vspace{0.7cm}
\includegraphics[scale=0.465,angle=-90,trim=10mm 0mm 0mm 0mm]{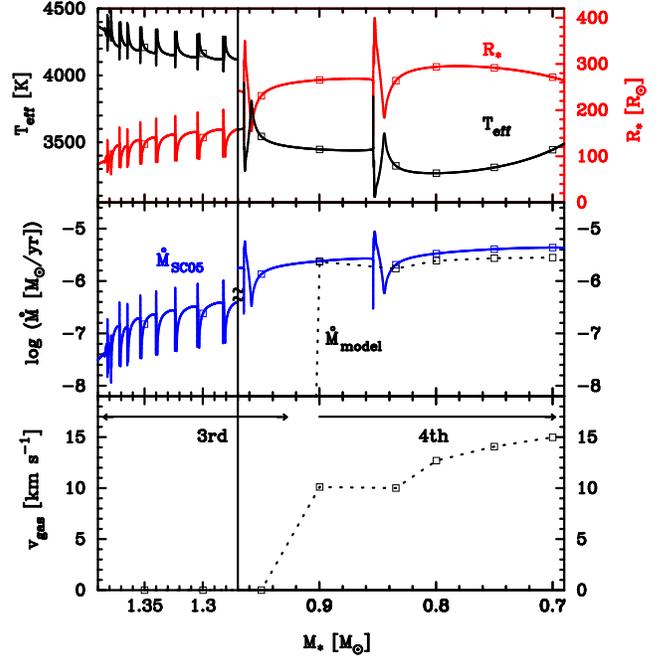}
\vspace{0.7cm}
\caption{Same as Figure~\ref{fig1}, but for the TP-AGB.
(A color version of this figure is available in the online journal.)\label{fig5}}
\end{figure}

Figure~\ref{fig5} shows the time evolution of the stellar parameters and the wind properties on the TP-AGB. 
The Alfv\'en wave-driven wind again starts to blow continuously after $M_{\ast}$ $\leq$ 0.8998 
$M_{\odot}$ (see the middle and bottom panels), firstly because of the decrease of the surface gravity 
and secondly because of the suppression of the damping of the magnetic energy associated with the 
Alfv\'en waves (see Section 4.1 for the details). We classify this wind as the four type which is 
characterized as a dense and slow stable wind appearing in the late stage of TP-AGB. The 
calculated values of $\dot{M}$ and $v_{\rm gas}$ are 1.7--2.8 $\times$ 10$^{-6} M_{\odot}$ yr$^{-1}$ 
and 10--15 km s$^{-1}$, respectively.

\begin{figure}
\vspace{0.7cm}
\includegraphics[scale=0.42,angle=-90,trim=10mm 0mm 0mm 0mm]{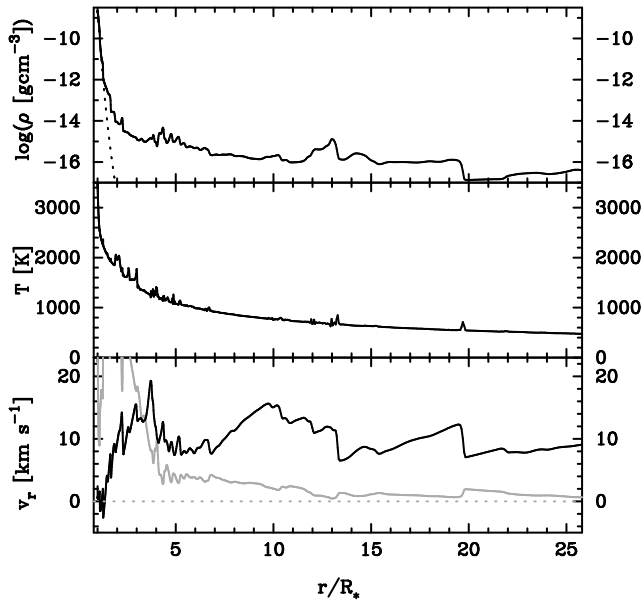}
\vspace{0.7cm}
\caption{Same as Figure~\ref{fig3}, but at $M_{\ast}$ = 0.8998 $M_{\odot}$ at simulation time $t$ = 96.2 yr, 
which is a typical wind of the fourth type.\label{fig6}}
\end{figure}

A snapshot of the radial structure at $M_{\ast}$ = 0.8998 $M_{\odot}$ on the TP-AGB is shown in 
Figure~\ref{fig6}. We can see from the middle panel that the chromospheric structure 
does not appear.  Nevertheless the stable outflow streams out 
due to the low surface gravity of $\log g = -0.455$. The temperature profile is almost smooth except for 
the regions at $r=2.0$, $2.6$, and $3.0$ $R_{\ast}$ because the gas temperature is primarily determined by 
the radiation on the TP-AGB; the temperature of gas heated in the high density flow rapidly 
falls down to the radiative equilibrium one. Furthermore, in the region outside 4.0 $R_{\ast}$,  the wind is super-Alfv\'enic, 
$v_{r}>v_{\rm A}$ (see the bottom panel), which is in sharp contrast to the first and second types of  
wind shown in Figures~\ref{fig2} and~\ref{fig3}. Thus the physical properties of the fourth type wind is 
quite different from those of the first or second type winds.

The presence of the fourth type wind clearly demonstrates that Alfv\'en wave-driven mechanism is promising 
even for TP-AGB stars not emitting chromospherically active lines. Also, it should be pointed out that the 
$M_{\rm ini}$ = 1.5 $M_{\odot}$ model star is oxygen-rich before the onset of final TP at 
$M_{\ast}$ = 0.8539 $M_{\odot}$. Thus the Alfv\'en wave-driven mechanism should be considered to be one 
of possible mechanism to trigger the mass loss from O-rich AGB stars for which no self-consistent hydrodynamical 
model of pulsation-enhanced dust-driven wind is available (see Section 1). 

\subsubsection{$M_{\rm ini}$ = 2.0 $M_{\odot}$ model} 
\begin{figure}
\vspace{0.7cm}
\includegraphics[scale=0.465,angle=-90,trim=10mm 0mm 0mm 0mm]{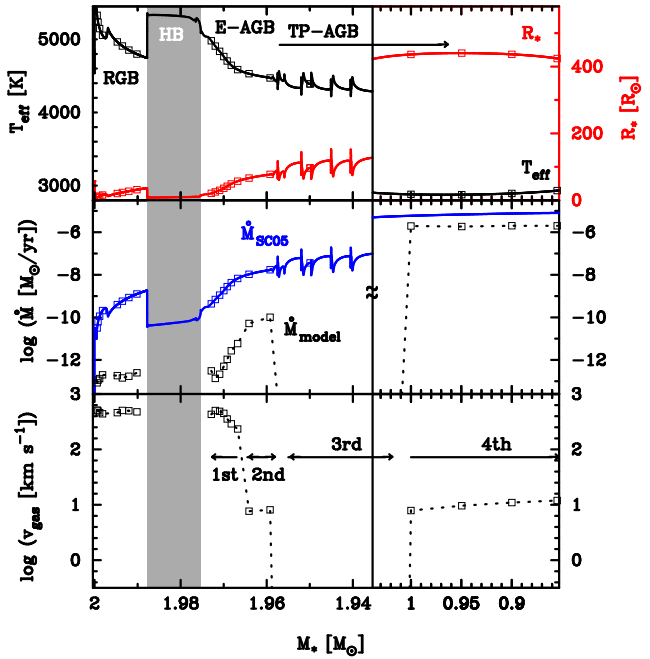}
\vspace{0.7cm}
\caption{Same as Figure~\ref{fig1}, but for the model star of $M_{\rm ini}$ = 2.0 $M_{\odot}$ 
and the abscissa ranging from the RGB to the TP-AGB phase.
%Same as Figure~\ref{fig7}, but the abscissa ranging from the RGB to the TP-AGB phase.
(A color version of this figure is available in the online journal.)\label{fig7}}
\end{figure}

Figure~\ref{fig7}  shows the time evolution of the stellar parameters (top panel) and the wind properties 
(middle and bottom panels) from the RGB to the TP-AGB phase. 
Contrary to the model star of $M_{\rm ini}$ = 1.5 $M_{\odot}$, stable outflows (the first and second 
type winds) are driven in all the simulated stars sampled from the  RGB to the E-AGB phase. 
The transition from the first type to the second type occurs at $M_{\ast}$ = 1.964 $M_{\odot}$ 
($R_{\ast}$ = 66.36 $R_{\odot}$).  
%at which both types appear alternately. 
The wind types changes to the third 
type at $M_{\ast}$ = 1.950 $M_{\odot}$ ($R_{\ast}$ = 96.50 $R_{\odot}$) when the model star enters into 
the TP-AGB phase.

The fourth type wind onsets at $M_{\ast}$ = 1.000 $M_{\odot}$. 
In the evolutionary stage after 1.000 $M_{\odot}$, the star is in the C-rich AGB phase.  
During this phase $\dot{M}$ = 1.8--2.0 $\times$ 10$^{-6} M_{\odot}$ yr$^{-1}$ and 
$v_{\rm gas}$ = 7.9--12 km s$^{-1}$, respectively.  
Although $v_{\rm gas}$ increases with decreasing the stellar mass $M_{\ast}$, 
the mass-loss rate $\dot{M}$ does not increase, in contrast to the model of $M_{\rm ini}$ = 1.5 $M_{\odot}$. 

\subsubsection{$M_{\rm ini}$ = 3.0 $M_{\odot}$ model} 

\begin{figure}
\vspace{0.7cm}
\includegraphics[scale=0.465,angle=-90,trim=10mm 0mm 0mm 0mm]{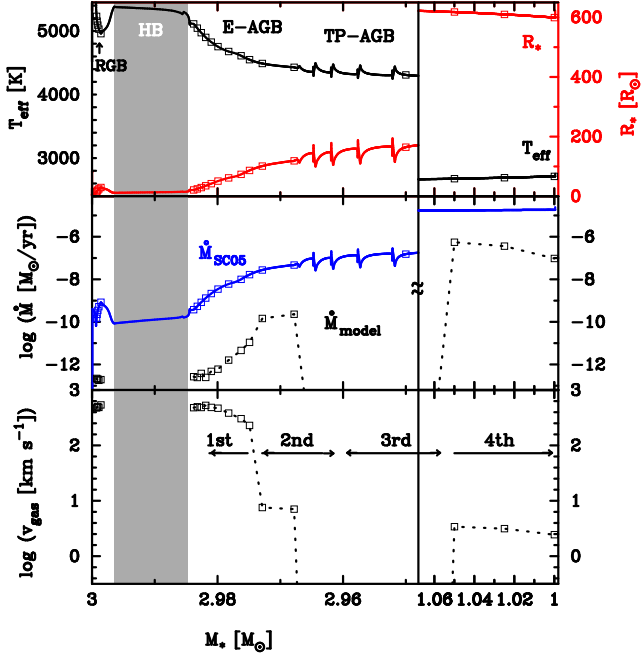}
\vspace{0.7cm}
\caption{Same as Figure~\ref{fig1}, but for the model star of $M_{\rm ini}$ = 3.0 $M_{\odot}$ 
and the abscissa ranging from the RGB to the TP-AGB phase.
%\caption{Same as Figure 9, but the abscissa ranging from the RGB to the TP-AGB phase.
(A color version of this figure is available in the online journal.)\label{fig8}}
\end{figure}

The evolution of the stellar parameters and the wind properties from the RGB to AGB is depicted in Figure~\ref{fig8}. 
The transition from the first type to the second type occurs at $M_{\ast}$ = 2.973 $M_{\odot}$ 
($R_{\ast}$ = 101.1 $R_{\odot}$). The wind types changes to the third 
type at $M_{\ast}$ = 2.950 $M_{\odot}$ ($R_{\ast}$ = 164.6 $R_{\odot}$) when the model star enters into 
the TP-AGB phase, as in the model of $M_{\rm ini}$ = 2.0 $M_{\odot}$.

%We can see from the bottom panel that the model star does not develop the second type wind during the 
%evolution, being different from the previous two model stars; only first type wind appears on the RGB, and the 
%transition from the first type to the third type occurs between at $M_{\ast}$ = 2.976 $M_{\odot}$ 
%($R_{\ast}$ = 73.37 $R_{\odot}$) and $M_{\ast}$ = 2.975 $M_{\odot}$ ($R_{\ast}$ = 86.43 $R_{\odot}$) on the E-AGB. 

%enters into the TP-AGB phase at $M_{\ast}$ = 2.965 $M_{\odot}$, and turns to C-rich AGB stars 

Then the model star turns to C-rich AGB stars
at  $M_{\ast}$ = 2.186 $M_{\odot}$. During the evolution, the effective temperature $T_{\rm eff}$ in the interpulse phase 
decreases and then turn up. After reaching to the minimum ($\sim$ 2620 K), the fourth type appears at 
$M_{\ast}$ = 1.050 $M_{\odot}$ with $\dot{M}$ = 5.5 $\times$ 10$^{-7} M_{\odot}$ yr$^{-1}$ and 
$v_{\rm gas}$ = 3.4 km s$^{-1}$ (Figure~\ref{fig8}). However, in contrast to the previous two models, $\dot{M}$ 
and $v_{\rm gas}$ gradually decrease as the star evolves; $\dot{M}$ = 3.6 (0.95) 
$\times$ 10$^{-7} M_{\odot}$ yr$^{-1}$  and $v_{\rm gas}$ = 3.1 (2.4) km s$^{-1}$ at $M_{\ast}$ = 1.025 (1.000) 
$M_{\odot}$. The behavior of the fourth type wind is supposed to reflect the lower gas density in the surface 
layer as well as the smaller surface gravity of the model star, in comparison with stars with smaller 
$M_{\rm ini}$; the model star is more luminous and undergoes more frequent TDUs which lead to more abundant 
carbon and larger $\kappa_{\rm surf}$ in the surface layer and lower $T_{\rm eff}$.

\subsection{Transition of wind type}

\begin{figure}
\vspace{0.7cm}
\includegraphics[scale=0.38, angle=-90,trim=0mm 0mm 0mm 0mm]{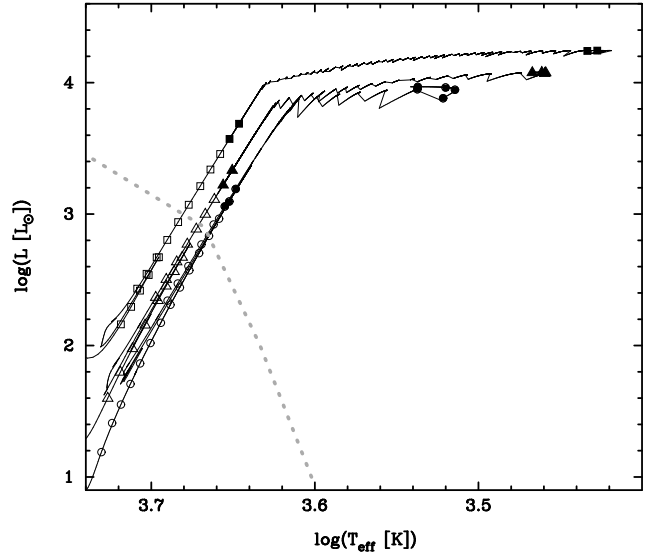}
\vspace{0.7cm}
\caption{The position of simulated stars exhibiting the stable wind on the evolution 
tracks of the model stars (thin lines) in the HR diagram; the circles, triangles, and squares denote 
the simulated stars with $M_{\rm ini}$ = 1.5, 2.0, and 3.0 $M_{\odot}$, respectively. 
The open (filled) symbols indicate $v_{{\rm gas}}>$ ($<$) 80 km s$^{-1}$.
For a reference, the dotted line shows the dividing line taken from Figure 2 of \citet{lin79}. 
Note that we delete the tracks during the periods of thermal pulse to avoid overlapping the tracks. 
\label{fig9}}
\end{figure}

Figure~\ref{fig9} plots the position of simulated stars exhibiting the stable wind on the evolution 
tracks of the model stars (thin lines) in the HR diagram; the circles, triangles, and squares denote 
the simulated stars with $M_{\rm ini}$ = 1.5, 2.0, and 3.0 $M_{\odot}$, respectively. 
The open (filled) symbols indicate $v_{{\rm gas}}>$ ($<$) 80 km s$^{-1}$.

We can see from the figure that the transition from the first type to second type seems to depend only 
on the effective temperature; the warm and fast wind (the first type wind) can be seen only in the region 
of  $\log T_{\rm eff}$ $\geq$ 3.66. Also, as can be seen Figure~\ref{fig9}, the star with the second type 
wind characterized by $v_{{\rm gas}} < 10$ km s$^{-1}$, $\dot{M}< 10^{-10} M_{\odot}$ yr$^{-1}$, and 
relatively cool wind (Figures~\ref{fig2} and~\ref{fig3}) are located in the very narrow range of 
3.646 $<$ $\log T_{\rm eff}$ $<$ 3.656.  
On the other hand, the fourth type (slow and massive) wind can 
be seen in the late stage of TP-AGB, and the position of transition to the fourth type depends on not 
only the effective temperature but also the stellar luminosity. The behavior of the fourth type wind 
seems to depend on the surface gravity as is mentioned in Section 3.2.1.  
In this section, we shall investigate how the wind properties and the transition of wind type depend on 
the effective temperature and the surface gravity of stars.

\subsubsection{Dependence on effective temperature}

\begin{figure}
\vspace{0.7cm}
\includegraphics[scale=0.34,angle=-90,trim=10mm 0mm 0mm 0mm]{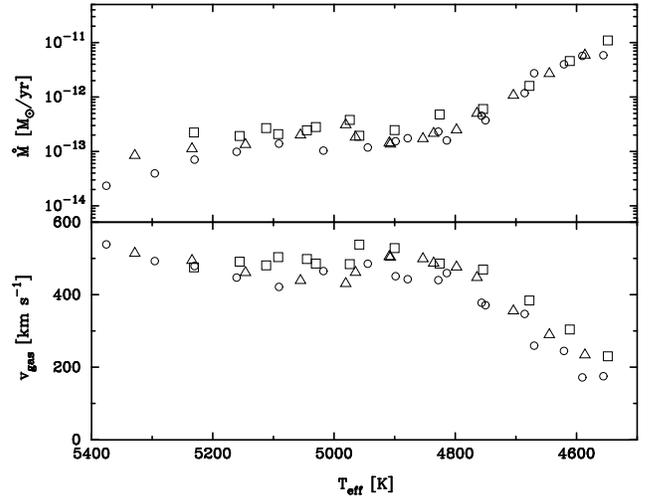}
\vspace{0.7cm}
\caption{$\dot{M}$ (top) and $v_{{\rm gas}}$ (bottom) versus $T_{\rm eff}$ for the first type wind. 
Circles, triangles, and squares denote results of for the simulated stars of $M_{\rm ini}$ = 1.5, 2.0, and 3.0 $M_{\odot}$, 
respectively.\label{fig10}}
\end{figure}

Figure~\ref{fig10} shows the dependence of the mass-loss rate $\dot{M}$ (top) and outflow velocity 
$v_{{\rm gas}}$ (bottom) on the effective temperature $T_{\rm eff}$ for the the first type wind. 
Roughly speaking, there is a tendency that $\dot{M}$ increases and $v_{{\rm gas}}$ decreases 
with decreasing $T_{\rm eff}$  except for the region where 4800 K $< $$T_{\rm eff}$ $<$ 5200 K.

%Roughly speaking, $\dot{M}$ tends to increase with decreasing $T_{\rm eff}$. The dependence of $v_{{\rm gas}}$ on 
%$T_{\rm eff}$ differs in the following three temperature region. In the 
%temperature region of $T_{\rm eff}$ $>$ 5250 K, $v_{{\rm gas}}$ somewhat decreases 
%with decreasing $T_{\rm eff}$, which reflects the increase of the fluctuation amplitude 
%$\langle\delta v_{{\rm 0}}\rangle$ as mentioned in Section 3.2.1. Then, in the temperature region 
%where 4700 K $< $$T_{\rm eff}$ $<$ 5250 K,  $v_{{\rm gas}}$ 
%tends to increase with decreasing $T_{\rm eff}$. This reflects the changes of the position of 
%and the gas density at the base of wind as later mentioned in Section 3.3.2. 
%In the temperature region of $T_{\rm eff}$ $\la$ 4700 K,   $v_{{\rm gas}}$ decreases with 
%decreasing $T_{\rm eff}$ for the simulated stars with $M_{\rm ini}$ = 1.5 and 2.0 $M_{\odot}$, 
%which could indicate that the transition from the first type to the second type is gradual. 
 
%As can be sen Figure~\ref{fig9}, the star with the second type wind characterized by 
%$v_{{\rm gas}} < 10$ km s$^{-1}$, $\dot{M}< 10^{-10} M_{\odot}$ yr$^{-1}$, and relatively cool wind 
%(Figures~\ref{fig2} and~\ref{fig3}) are located in the very narrow range of 
%\textbf{3.646} $<$ $\log T_{\rm eff}$ $<$ \textbf{3.656}.  

\begin{figure}
\vspace{0.7cm}
\includegraphics[scale=0.34,angle=-90,trim=10mm 0mm 0mm 0mm]{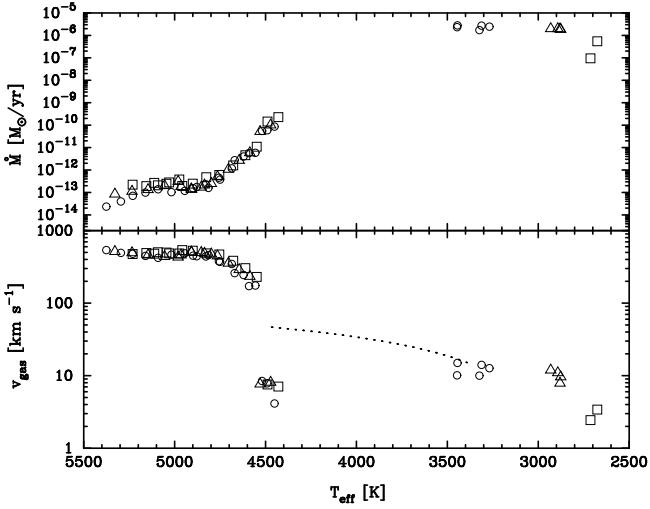}
\vspace{0.7cm}
\caption{Same as Figure 10, but for the range from the RGB to the TP-AGB phase 
with the addition of the slow and dense winds, and equation (2) of \citet{woo16} 
(dotted line; see Section 4.2.1).\label{fig11}}
\end{figure}

Figure~\ref{fig11} plots $\dot{M}$ (top) and $v_{{\rm gas}}$ (bottom) for all the simulated stars with 
the stable wind on the RGB and the AGB. 
The figure clearly shows that the 
transition from the first to second type occurs at $T_{\rm eff}$ = 4530 K, and the the second type 
wind disappears at  $T_{\rm eff} = 4420$ K. The stable wind is never driven from the simulated stars 
whose effective temperature are in the range of $3450$ K $<T_{\rm eff} < 4420$ K,  
although the atmospheres are lifted up by the  Alfv\'en waves (Figure~\ref{fig4}). 
In $ T_{\rm eff} < 3450$ K , dense, slow, and cool stable winds develop by the 
dissipation of long wavelength Alfv\'en waves in the wind acceleration region (see Section 4.1).

\subsubsection{Dependence on surface gravity}

\begin{figure}
\vspace{0.7cm}
\includegraphics[scale=0.34,angle=-90,trim=10mm 0mm 0mm 0mm]{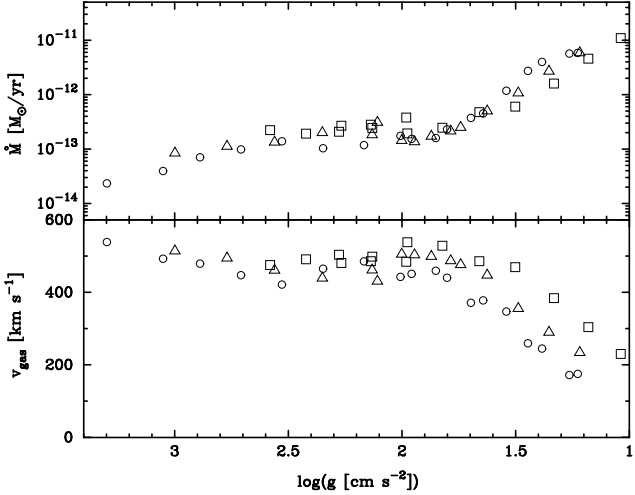}
\vspace{0.7cm}
\caption{Same as Figure~\ref{fig10} but plotted on $\log g$.\label{fig12}}
\end{figure}

Figure~\ref{fig12} plots $\dot{M}$ (top) and $v_{{\rm gas}}$ (bottom) versus $\log g$ for the first 
type wind. $\dot{M}$ increases and $v_{{\rm gas}}$ decreases with decreasing $\log g$
except for the region of 1.9 $<$ $\log g$ $<$ 2.5.

\begin{figure}
\vspace{0.7cm}
\includegraphics[scale=0.34,angle=-90,trim=10mm 0mm 0mm 0mm]{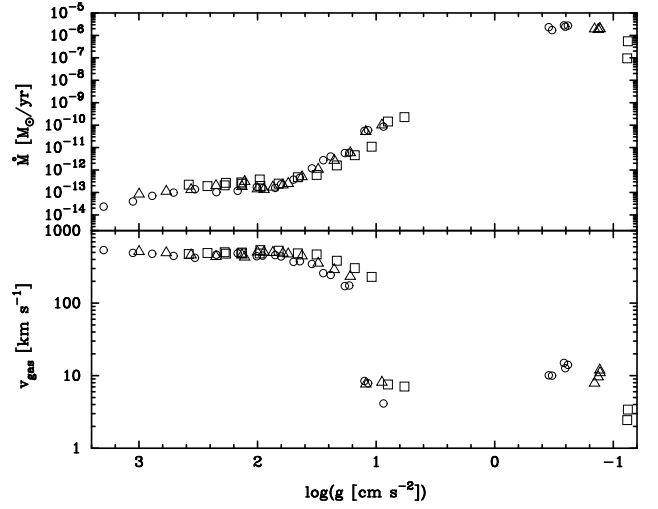}
\vspace{0.7cm}
\caption{Same as Figure 12, but from the RGB to the TP-AGB phase. 
\label{fig13}}
\end{figure}

%Only the $M_{\rm ini}$ = 1.5 and 2.0 $M_{\odot}$ model stars can develop the second type wind in 
%the range of $0.94 < \log g < 1.3$ as shown in Figure~\ref{fig13}. The stellar wind becomes second type 
%wind at $\log g$ = 1.218 (1.265) for the simualted star of $M_{\rm ini}$ = 1.5 (2.0) $M_{\odot}$.
As is presented in Figure~\ref{fig13}, the wind turns to 
the second type at $\log g$ = 1.098 (1.087, 0.901) for the simulated star of 
$M_{\rm ini}$ = 1.5 (2.0, 3.0) $M_{\odot}$.
No stable wind is driven in the range of 
%$-0.456<\log g <0.908$ 
-0.456 $<$ $\log g$ $<$0.475 
and the value of $\log g$ at 
which the transition to the unstable (the third type) wind occurs slightly depend on $M_{\rm ini}$;  
$\log g$ between 0.779 and 1.068, (0.759 and 0.951, 0.475 and 0.761)
for $M_{\rm ini}$ = 1.5, (2.0, 3.0) $M_{\odot}$, respectively. 

Although AGB stars can develop the fourth type wind below $\log g$ = -0.456 in the model stars considered in 
this paper, the value of  $\log g$  at which the fourth type wind starts to blow significantly depends on 
$M_{\rm ini}$. Also the wind properties during the evolution are influenced by not only the surface gravity 
but also the surface opacity $\kappa_{{\rm surf}}$, as mentioned in Section 3.2.3. The condition for the 
apperance of the fourth type wind derived from the examination of the results of MHD simulations 
is expressed as 
\begin{eqnarray}
\log \left[\frac{g}{\text{cm s$^{-2}$}}\left(\frac{\kappa_{{\rm surf}}}{\text{cm$^{2}$g$^{-1}$}}\right)^{1/2}
\right] \leq -2.1.\label{revival_condition}
\end{eqnarray}

\section{Discussions}

\subsection{Analysis of wind properties}

Here we examine how the wind types found in the MHD simulations relate with the dissipation of Alfv\'en waves, 
using the time averaged radial profiles of physical quantities characterizing the Alfv\'en wave-driven wind. 
The radial variation of magnetic perturbation being associated with the dissipation of Alfv\'en waves 
directly affect the properties of stellar winds, which is expressed in equation~(\ref{gas_ene_eq}). 
We introduce the relative variation $X$ at $r$ defined as 
\begin{eqnarray}
X(r)\equiv\frac{r^{2}f(B_{\perp,1}^{2}+B_{\perp,2}^{2})}{R_{\ast}^{2}f(R_{\ast})(B_{\perp,\ast,1}^{2}
+B_{\perp,\ast,2}^{2})}\label{damp_eq}
\end{eqnarray} 
where $B_{\perp,\ast,1}$ ($B_{\perp,\ast,2}$) is the first (second) tangential 
component of the magnetic field strength at the stellar surface. The actual tangential components of these quantities 
at the stellar surface are set to be zero as the boundary condition while the tangential components of gas 
velocity at the radial position are set as the input parameters as described in Section 2.3. 
We substitute the values at the stellar surface with the actual values 
at the second innermost radial computational grid. The value of the magnetic energy density derived from the substituted 
quantities  $(B_{\perp,\ast,1}^{2}+B_{\perp,\ast,2}^{2})/(8\pi)$ is in the range of 0.2 to 0.8 $\times$ 
$\rho\langle\delta v_{{\rm 0}}\rangle^{2}$.

\begin{figure}
\vspace{0.7cm}
\includegraphics[scale=0.55,angle=-90,trim=10mm 0mm 0mm 0mm]{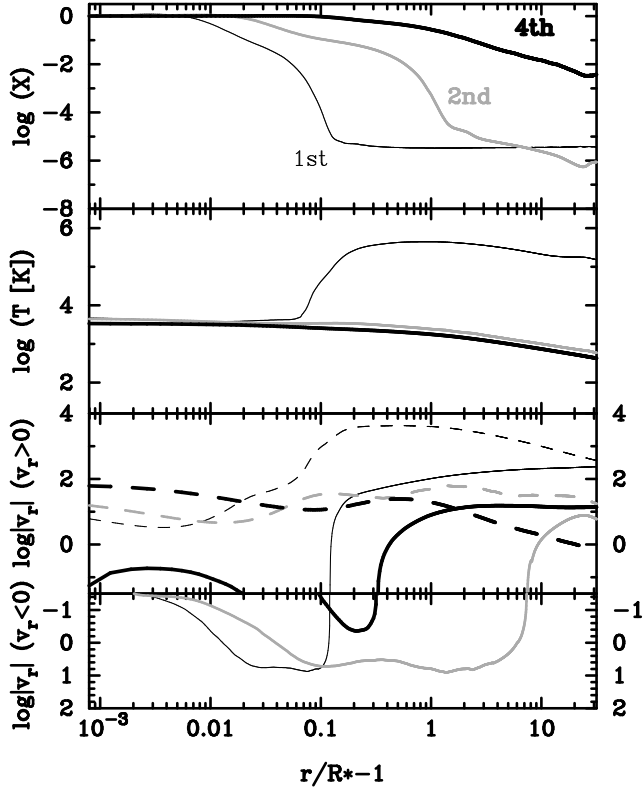}
\vspace{0.7cm}
\caption{The time averaged radial structures of the quantity X (top), the gas temperature $T$ (second), 
gas velocity $v_{r}$ (solid) and the Alfv\'en velocity $v_{\rm A}$ (dashed) (third and fourth).  
Thin black, gray, and thick black lines for the radial structures at 
$M_{\ast}$ = 1.4918, 1.4009, and 0.8998 $M_{\odot}$, respectively. \label{fig14}}
\end{figure}

Hereafter we investigate the relation between the wind types and the time averaged value of $X$ 
for the model star of  $M_{\rm ini}$ = 1.5 $M_{\odot}$. 
The radial structures of the time averaged values of $X$ are shown in the top panel in Figure~\ref{fig14} 
together with the time-averaged radial structures of the  gas temperature, the radial velocity, $v_{r}$, and  
the Alfv\'en velocity, $v_{\rm A}$, in the panels from the second to bottom; Thin black, gray, and thick black 
lines represent the typical radial structures for the first type wind at $M_{\ast}$ = 1.4918 $M_{\odot}$, the 
second type wind at $M_{\ast}$ = 1.4009 $M_{\odot}$, and the fourth type wind at $M_{\ast}$ = 0.8998 $M_{\odot}$, 
respectively. 

In the first type wind, $X$ decreases in a exponential manner according to the density structure in the 
inner nearly static region, and is not so changed in the outer wind region. 
The value of $X$ decreases by more than five orders of magnitude from the stellar surface to the outer wind region. 
Even at the radial position where the distance from the star $r$ - $R_{\ast}$ is several $\times$ 10$^{-2}$ $R_{\ast}$,  
$X$ decreases by two orders of magnitude and the gas temperature begins to increase due to the dissipation 
of Alfv\'en waves. Then the gas velocity starts to increase steeply around $r$ - $R_{\ast}$ is 0.1--0.2 
$R_{\ast}$. The increase of the gas temperature decreases the relative abundances of neutral species, 
which suppresses the ambipolar diffusion and the further decrease of $X$ in the outer wind part.

%As the star swells up with the evolution, the input of 
%the magnetic energy at the stellar surface changes; it slightly increases with $\langle\delta v_{{\rm 0}}\rangle$
%for the star with $T_{\rm eff}$ $>$  5250 K ($\log g$ $>$ 2.8), which results in smaller outflow velocity $v_{{\rm gas}}$ 
%in denser wind; then it decreases by one order of magnitude because of the decrease of the density for the star with 
%$T_{\rm eff}$ $<$  5250 K ($\log g$ $<$ 2.8). However, due to the 
%increase of the pressure scale height normalized by stellar radius at the stellar surface $h_{{\rm p}}/R_{\ast}$,  
%$X$ decreases much more gradually with the distance from the stellar surface $r$ - $R_{\ast}$ 
%(see the top panel of Figure~\ref{fig14}). As a result, the magnetic 
%energy density in the region relatively far from the stellar surface becomes higher in the star with larger radius. 
%Then the dissipation of Alfv\'en waves for the star occurs at a larger distance in unit of $(r-R_{\ast})/R_{\ast}$. 
%Thus a wind launches from a higher altitude, $(r-R_{\ast})/R_{\ast}$ with the evolution. 
%For the star with the stellar parameter of 4700 K $<$ $T_{\rm eff}$ $<$ 5250 K 
%(1.5 $<$ $\log g$ $<$ 2.8), $v_{{\rm gas}}$ of the first type wind increases with the slight decrease 
%of the gas density in the wind base for the star as mentioned in Section 3.3.2. 

After the effective temperature $T_{\rm eff}$ drops down below 4900 K, as the star evolves, the wind 
speed of the first type wind decreases the with stellar evolution. This seems to be ascribed to 
the decrease of the thermal enthalpy due to the more efficient radiative cooling in denser wind. 
Then the winds become denser and slower (second type) as long as a part of the dissipated 
energy is converted to the kinetic energy of gas necessary to maintain the stable outflow. Considerably 
later the stable wind in the fourth type is formed. As shown in the second panel in Figure~\ref{fig14}, 
the temperature at the wind base in this fourth type is quite lower than that for the first type. The 
determination of the stability of the various types of winds 
requires the analysis of the energetics in the outer wind region.

\begin{figure}
\vspace{0.7cm}
\includegraphics[scale=0.55,angle=-90,trim=10mm 0mm 0mm 0mm]{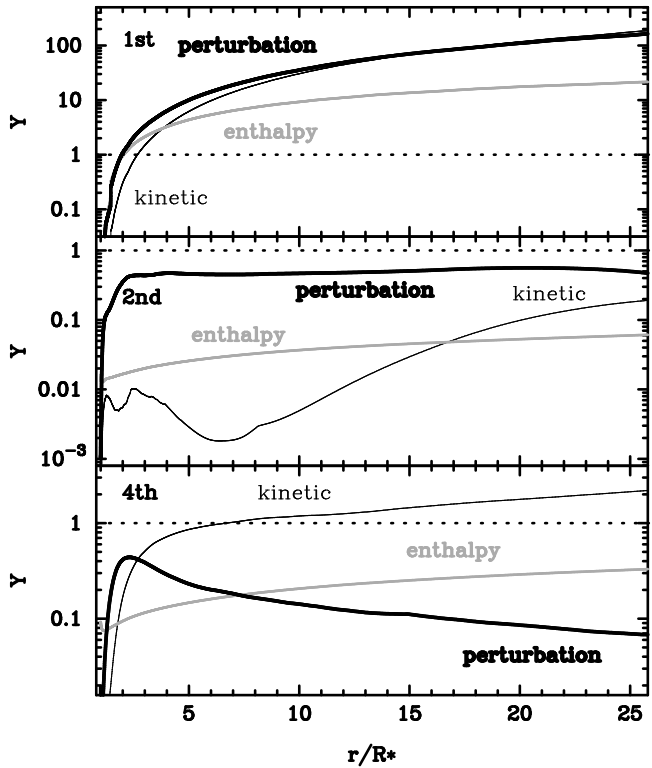}
\vspace{1.3cm}
\caption{The radial structure of the energy densities normalized by the absolute value of the product of 
the gas density and the gravitational potential energy $Y$-s. Top, middle, and bottom panel for the 
results at $M_{\ast}$ = 1.4918, 1.4009, and 0.8998 $M_{\odot}$, 
respectively. Thin black, gray, and thick black lines for the kinetic energy ($Y_{\rm 1}$), thermal enthalpy 
($Y_{\rm 2}$), and perturbation energy ($Y_{\rm 3}$), respectively. Dashed line for the gravitational 
potential. \label{fig15}}
\end{figure}

We analyze the energetics in the outer wind region by introducing the quantities $Y_{i}$ ($i$=1,2,3) defined as 
\begin{eqnarray}
& & Y_{\rm 1}=\frac{\rho v_{r}^{2}/2}{\rho GM_{\ast}/r},\label{def_Y1}\\
& & Y_{\rm 2}=\frac{\gamma p/(\gamma-1)}{\rho GM_{\ast}/r}\label{def_Y2},\\
& & Y_{\rm 3}=\frac{\rho(v_{\perp,1}^{2}+v_{\perp,2}^{2})/2+(B_{\perp,1}^{2}+B_{\perp,2}^{2})/(8\pi)}
{\rho GM_{\ast}/r},\label{def_Y3}
\end{eqnarray}
where $Y_{\rm 1}$, $Y_{\rm 2}$, and $Y_{\rm 3}$ are kinetic energy density, thermal enthalpy, and the magnetic energy 
density associated with the Alfv\'en waves normalized by the absolute value of 
the product of the gas density and the gravitational potential energy, respectively. We derive these quantities from 
the time averages of numerators and denominators on the right-hand sides of 
equations~(\ref{def_Y1}), (\ref{def_Y2}), and 
(\ref{def_Y3}). The radial profiles of $Y$-s are given in Figure~\ref{fig15}. Here we note that 
unnormalized energy densities themselves decrease with $r$. 

The top panel of Figure~\ref{fig15} shows the results at $M_{\ast}$ = 1.4918 $M_{\odot}$. In this first type wind, 
the three kinds of energy are predominant over the gravitational energy in the outer wind part, and the 
wind is warm and fast. The predominance of thermal enthalpy leads to the suppression of the ambipolar diffusion, which 
results in the predominance of the perturbation energy.

In the second type wind at  $M_{\ast}$ =  1.4009 $M_{\odot}$ (the middle panel of Figure~\ref{fig15}), 
$Y_{\rm 3}$ is much less than that in the first type wind, and is 0.4-0.6 in the region of 
$r$ $>$ 2.2 $R_{\ast}$.  
This is due to the damping of the magnetic perturbation in the outer wind region as 
shown in the top panel of Figure~\ref{fig14}. 
%However 
$Y_{\rm 2}$ is less than 0.1 through the outer wind region because of the more efficient radiative 
cooling in denser wind. Also, in this type, the gas temperature is not strongly affected by the dissipation of Alfv\'en waves; 
as shown in Figure~\ref{fig3}, the gas temperature does not deviate from the radiative equilibrium one so much. 
%except in the region at $r$ = 1.58 $R_{\ast}$ where the gas is heated up by the accretion shock. 
Therefore, although the perturbation energy is not so quite lower than the gravitational energy, the wind is slow 
due to the inefficient energy conversion from the magnetic energy to the kinetic energy. 
The value of $Y_{\rm 1}$ never exceeds unity even at the outer boundary (25$R_{\ast}$); 
This casts doubt on the stability of the second type wind. It is considered that the sporadic event such as the 
propagation of shock barely enables the second type wind to stream out continuously. 

The results at $M_{\ast}$ = 0.8998 $M_{\odot}$ for the fourth type wind is presented in the bottom panel of 
Figure~\ref{fig15}. We can see that $Y_{\rm 3}$ is 0.07-0.4 in the region of $r$ $>$ 2.0 $R_{\ast}$; it reaches the 
maximum value (0.4) at $r$ = 2.3 $R_{\ast}$ and decrease with $r$. 
This attributes to the further suppression of the damping of the magnetic perturbation in the inner 
atmosphere (see top panel of Figure~\ref{fig14}), and the efficient energy conversion in the wind part.  
In the region of $r$ $>$ 2.3 $R_{\ast}$, $Y_{\rm 2}$ is larger than 0.1, and is larger than that in the second type wind, 
which reflects the considerable decrease of the surface gravity arising from the increase of $R_{\ast}$ and the 
decrease of $M_{\ast}$ although $T_{\rm eff}$ substantially decreases. Also, as in the second type, the gas temperature 
is not strongly affected by the dissipation of Alfv\'en waves as shown in Figure~\ref{fig6}. However the ratio of the 
thermal enthalpy to the perturbation energy ($Y_{\rm 2}$/$Y_{\rm 3}$) is higher, and the energy conversion from the 
perturbation to kinetic energy is more efficient than in the second type. The kinetic energy increases with 
increasing the distance, while the perturbation energy decreases. 
At $r$ $=$ 6.7 $R_{\ast}$, $Y_{\rm 1}$ reaches 1.0, and the stable super-Alfv\'enic wind is formed. 

The results of analysis are summarized as follows: The transition from the first to second type occurs mainly by the 
suppression of the thermal enthalpy at the base of the wind.  The second type wind could not be always stable. The 
appearance of the fourth type wind seems to be related with the degree of the suppression of the damping of 
the magnetic perturbation in the inner region; a larger 
fraction of the input Alfv\'enic Poynting flux remains at a higher altitude to derive dense wind. 

\subsection{Comparison with observations} 

\subsubsection{Comparison with RGB and E-AGB stars} 

The mass-loss rate $\dot{M}$ of the first type wind seen in the simulated RGB and E-AGB stars 
with $T_{\rm eff}$ ($\log g$) higher than 4530 K (1.1) is at least two orders of magnitude smaller than 
the rates from the empirical formula of \citet{sch05} (see middle panels of 
Figures~\ref{fig1}, ~\ref{fig7} and~\ref{fig8}); The calculated $\dot{M}$ is also smaller than 
the rates of metal poor red giants with high $T_{\rm eff}$ evaluated from the optical/near infrared observations  
by \citet{dup09} and \citet{mes09}, despite that the mass-loss rate is considered to be higher 
in metal--rich star than metal--poor star \citet{mes09} or little dependent of metallicity \citet{mcd15}.

The outflow velocity $v_{{\rm gas}}$ of the simulated stars with high $T_{\rm eff}$ ($\ga$ 4530 K) 
exceeds 200 km s$^{-1}$. The value of $v_{{\rm gas}}$ is hard to estimate from observations of 
chromospheric lines while the observations of He I 10830 $\rm{\AA}$  line by \citet{dup09} may provide the radial 
velocity of the wind with gas temperature higher than 20000 K. However, for 13 giants with high $T_{\rm eff}$ 
in their samples, $v_{{\rm gas}}$ is at most 104 km s$^{-1}$ for HD121135. The difference of $v_{{\rm gas}}$ 
may reflect, at least in part, that their estimated values do not necessarily represent the wind velocities 
in outer low density region; As shown in the bottom panel of Figure~\ref{fig2}, the outflowing gas increases with 
increasing the distance from the center, even at $\sim$ 10 $R_{\ast}$, and it is not clear whether the higher 
velocity component contributes to the blue wing of observed line profiles or not.

The outflow velocity $v_{{\rm gas}}$ drops down lower than 10 km s$^{-1}$ at $T_{\rm eff}$ ($\log g$) 
around 4500 K (1.1) in our simulations, being accompanied with the transition from the first to second type. However, 
the UV observations of chromospheric lines clearly show that the wind velocities slow down lower than 20 
km s$^{-1}$ at $T_{\rm eff}$ = 3500 K more gradually, as is depicted in Figure \ref{fig11} by 
dotted line (see equation (2) of \citealt{woo16}). 

Thus, the problematic discrepancies between the results of MHD simulations and the observations 
are the following two. (1) the mass-loss rate for the stars with $T_{\rm eff}$ $\ga$ 4530 K. (2) the 
outflow velocity for the stars with $T_{\rm eff}$ in the range of 3450 to 4420 K. 

The first discrepancy may arise from low values of the input parameters  $f_{{\rm 0}}B_{r,{\rm 0}}$ and 
$\langle\delta v_{{\rm 0}}\rangle$; $f_{{\rm 0}}B_{r,{\rm 0}}$ being increased from 1 to 5G, $\dot{M}$ 
%tends to increase by an order of magnitude or more, 
increases by an order of magnitude or more in some cases.
%and the coronal wind with maximum temperature higher than 10$^{6}$ K often forms. 
Some active giants with high $T_{\rm eff}$ at the base of the RGB (EK Eri, V390 Aur, EI CnC, and 
$\iota$ Cap) have further strong surface magnetic field \citep{aur15}. We will investigate systematically the 
dependence of the surface magnetic field in future work. For the weakly-active magnetic star at the 
base of RGB such as Pollux whose surface averaged longitudinal magnitude is sub-Gauss \citep{aur09}, 
larger $\langle\delta v_{{\rm 0}}\rangle$ may be required; recently \citet{ogo17} estimates 
an upper limit of $\dot{M}$ of Pollux; 
$\dot{M}$ to be 3.7 $\times$ 10$^{-11} M_{\odot}$ yr$^{-1}$, which is higher than those of 
the first type wind in our model with $f_{{\rm 0}}B_{r,{\rm 0}}$ = 1G.  

On the other hand, the second discrepancy is not 
improved by adjusting the input parameters; When $f_{{\rm 0}}B_{r,{\rm 0}}$ increases from 1 to 5G, 
the second type wind changes to the third type wind due to the more efficient cooling in denser wind. 
In order to maintain the chromospheric structure and the continuous gas-outflow for 
the stars with $T_{\rm eff}$ in the range of 3450 to 4420 K  
%other additional heating mechanisms such as \textcolor{red}{the dissipation of acoustic 
%waves originating from the photospheric oscillation}; 
the formation of the local low density region(s) with high tempeature could be required,  which 
will be able to be investigated in higher dimensional simulations.  Also in this context, the 
applicability of the Rosseland mean opacity at the stellar surface should be examined since the 
radiative cooling rate evaluated in the present paper may be overestimated.

%Thus our MHD model cannot reproduce the observed wind properties of RGB and E-AGB stars with $T_{\rm eff}$ in the 
%range of 3500  to 4500 K even if the input parameters are changed. The wind from these stars 
%has been considered to be a cold wave-driven wind; \citet{cra11} estimated the mass-loss rate of cool stars theoretically 
%with the hot coronal one $\dot{M}_{{\rm hot}}$ and the cold wave-driven one $\dot{M}_{{\rm cold}}$ which follows 
%the development of \citet{hol83}. For these stars such as $\alpha$ Boo and 
%$\alpha$ Tau, the ratio $\dot{M}_{{\rm hot}}/\dot{M}_{{\rm cold}}$ is lower than 1.0, and their estimated values 
%differ from those estimated from observations only by an order of magnitude or less (see Tables 2 and 3 of 
%\citealt{cra11}). The irreproducibility in our simulations for these stars means that a cold wave-driven wind does 
%not stream out continuously as that in the steady--state model of \citet{hol83}. The stability of a cold wave-driven 
%wind should be investigated for more details in the future study.

\subsubsection{Comparison with TP-AGB stars} 

The ranges of mass-loss rate $\dot{M}$ and outflow velocity $v_{{\rm gas}}$ of the fourth type wind  
(see Table~\ref{tbl-1:classification}) reasonably cover the values observed in optically 
visible O-rich Mira ($\dot{M}$ = $6 \times$10$^{-9}$--$3 \times$ 10$^{-6} M_{\odot}$ yr$^{-1}$ and 
$v_{{\rm gas}}$ = $2.7$--$12$ km s$^{-1}$, \citealt{you95}) , carbon semiregular 
($2 \times $10$^{-8}$--$3 \times $ 10$^{-5} M_{\odot}$ yr$^{-1}$ and $4.0$--$28$  km s$^{-1}$, 
\citealt{ber05}), and carbon Mira variables ($1 \times $10$^{-7}$--$3 \times $ 10$^{-5} M_{\odot}$ yr$^{-1}$ 
and $7.2$--$22$  km s$^{-1}$, \citealt{ber05}). Therefore the Alfv\'en wave-driven mechanism is 
considered to be promising one to trigger and sustain the massive outflow observed in the late 
stage of TP-AGB stars. 

However, it is difficult to clarify whether the observed wind properties can be reproduced by the MHD 
simulation, since the evolutionary stage as well as the stellar parameters is not well constrained from 
the observations, in particular for dust--enshrouded AGB stars. For example, the effective temperature 
$T_{\rm eff}$ is significantly different, depending on the method of observation and analysis; $T_{\rm eff}$ of R Leo 
is set to be 1800 (2000) K in the radiative transfer models of  \citet{sch13} and 
\citet{ram14} while \citet{per04} have estimated $T_{\rm eff}$ to be 2980 K 
based on the interferometric observations.

Here we examine the properties of the fourth type wind  from the well-studied two TP-AGB stars; O-rich AGB star o Cet 
and C-rich AGB star IRC+10216. In Table~\ref{tbl-2:data_of_two_TP_AGB_stars}, 
the input parameters of the MHD simulations for the two TP-AGB stars are provided as well as the wind properties 
derived from the observations and the simulations. Among the input parameters, $M_{\ast}$, $L_{\ast}$, and 
$T_{{\rm eff}}$ are taken from the literatures (see the caption in Table~\ref{tbl-2:data_of_two_TP_AGB_stars}); 
the data are evaluated from the observations and the models. Since the surface elemental composition is not 
well-known for both stars, we set $\kappa_{{\rm surf}}$ to be 2.0 $\times$ 10$^{-4}$ cm$^{2}$g$^{-1}$ which 
is adopted in the hydrodynamical models of pulsation--enhanced dust--driven wind as a fiducial value 
\citep[e.g.,][]{bow88,fle92,yas12}. 

\begin{table}
\centering
\caption{Observed wind properties (mass-loss rate $\dot{M}_{\rm obs}$ and outflow velocity $v_{\rm gas,obs}$),  
input parameters of our wind models ($M_{\ast}$, $L_{\ast}$, $T_{{\rm eff}}$, $\kappa_{{\rm surf}}$, $f_{{\rm 0}}B_{r,{\rm 0}}$, 
and $\langle\delta v_{{\rm 0}}\rangle$)  and the wind properties derived from our model
(mass-loss rate $\dot{M}_{\rm model}$ and outflow velocity $v_{\rm gas,model}$) for two TP-AGB stars.
}\label{tbl-2:data_of_two_TP_AGB_stars}
\scalebox{0.86}{
\hspace{-13mm} 
%\hspace{-26mm} 
%\scalebox{0.7}{
%\scalebox{0.64}{
%\hspace{-37mm} 
\begin{tabular}{lcc}
\hline
Objects   & o Cet &  IRC+10216 \\
\hline
 $\dot{M}_{\rm obs}$ [$M_{\odot}$ yr$^{-1}$] & 4.4 $\times$ 10$^{-7}$(9.4 $\times$ 10$^{-8}$)$^{(1)}$ & 
 9.2 $\times$ 10$^{-6}$$^{(1)}$ \\ 
 & & 3.25 $\times$ 10$^{-5}$$^{(5)}$ \\ 
 $v_{\rm gas,obs}$ [km s$^{-1}$]  & 6.7(2.4)$^{(1)}$ & 14.6$^{(1)}$\\ 
 & & 14.0$^{(6)}$ \\ 
\hline
$M_{\ast}$ [$M_{\odot}$] & 1.18$^{(2)}$ & 0.8$^{(7)}$ \\
$L_{\ast}$ [$L_{\odot}$] ($R_{\ast}$ [$R_{\odot}$])& 8605$^{(3)}$(332)$^{(3)}$ & 
 8640$^{(8)}$(641.5)$^{(8,9)}$ \\
$T_{{\rm eff}}$ [K] & 3055$^{(3)}$ & 2200$^{(9)}$ \\
$\kappa_{{\rm surf}}$ [cm$^{2}$g$^{-1}$] & 2.0 $\times$ 10$^{-4}$ & 2.0 $\times$ 10$^{-4}$ \\
$f_{{\rm 0}}B_{r,{\rm 0}}$ [G] & 4.1$^{(4)}$ & 3.8$^{(6)}$ \\
$\langle\delta v_{{\rm 0}}\rangle$ [km s$^{-1}$] & 4.59 & 3.89 \\
\hline
 $\dot{M}_{\rm model}$ [$M_{\odot}$ yr$^{-1}$] & 1.07 $\times$ 10$^{-5}$ & 3.33 $\times$ 10$^{-5}$\\
 $v_{\rm gas,model}$ [km s$^{-1}$]& 19.4 & 11.4\\
\hline
\end{tabular}}

\vspace{-0.2cm}

\begin{flushleft}
References.(1)\citet{kna98};(2)\citet{wya83};(3)\citet{woo04};(4)\citet{thi13};(5)\citet{cro97};
(6)\citet{dut17};(7)\citet{lad10};(8)\citet{men12};(9)\citet{ive96}
\end{flushleft}
\end{table}

The observationally unknown value of $f_{{\rm 0}}B_{r,{\rm 0}}$ for o Cet is set to be 4.1G,  
by referring to the optimized value in the steady--state model of \citet{thi13}. Then, as shown in the second 
column of Table~\ref{tbl-2:data_of_two_TP_AGB_stars}, the cool and dense wind (the fourth type wind) can be 
generated by the Alfv\'en wave-driven mechanism. Thus this result demonstrates the formation of the micron--sized 
Fe--free silicate \citep{hof08} or Al$_{2}$O$_{3}$ core--silicate mantle dust \citep{koz97,koz98,hof16} is not 
crucial to induce the massive wind from O--rich AGB star. 

However, the simulated values of $\dot{M}$ and $v_{\rm gas}$ are substantially larger than those evaluated from the 
CO observations by \citet{kna98}; o Cet has double winds with $v_{\rm gas}$ of 2.4 and 6.7 km s$^{-1}$
($\dot{M}$ of 9.4 $\times$ 10$^{-8}$ and 4.4 $\times$ 10$^{-7} M_{\odot}$ yr$^{-1}$).

The observed slow and moderate density wind from o Cet may be reproduced by decreasing  $f_{{\rm 0}}B_{r,{\rm 0}}$ 
and $\langle\delta v_{{\rm 0}}\rangle$. Decreasing the values, we confirm that the wind is still stable as long as 
$\dot{M}$ is higher than 10$^{-6}$ $M_{\odot}$ yr$^{-1}$. 
However when $f_{{\rm 0}}B_{r,{\rm 0}}$ ($\langle\delta v_{{\rm 0}}\rangle$) is decreased to 1.5G 
(2.0 km s$^{-1}$), the wind becomes unstable: the observed stable wind with $\dot{M}$ around 
10$^{-7}$ $M_{\odot}$ yr$^{-1}$  is not reproduced in our model for o Cet.

Here we shall raise the following three possibilities to explain the discrepancy between the observation and our simulation 
for o Cet. (1) lower surface opacity, since larger opacity makes the wind slower and less dense, as is seen in the model 
star of $M_{\rm ini}$ = 3.0 $M_{\odot}$ (see Section 3.2.3). (2) the wind from o Cet is actually sporadic rather than stable; 
slow and less dense wind tends to be unstable not only in our MHD model but also in the pulsation--enhanced dust--driven 
wind model including the effect of drift of dust (\citealt{san08}; Yasuda et al. 2018, in preparation), and this 
possibility will be explored from the time variation of the radial velocity of H$_{2}$O masers 
(e.g., \citealt{sud17} for R Crt). (3) a hybrid mechanism including dust--driven wind mechanism which 
may stabilize the unstable wind (the third type wind) in our MHD model with lower $f_{{\rm 0}}B_{r,{\rm 0}}$ or 
lower $\langle\delta v_{{\rm 0}}\rangle$. The investigations of these possibilities are left for the future works.

The value of $f_{{\rm 0}}B_{r,{\rm 0}}$ for IRC+10216 is set to be 3.8G, by referring to \citet{dut17}. Among the stellar 
parameters, $M_{\ast}$ is set to 0.8 $M_{\odot}$ which is the intermediate value of the current core mass 
(0.7--0.9 $M_{\odot}$) estimated by \citet{lad10} since the core mass is actually lower expected from the adopted 
lower value of $L_{\ast}$. As can be seen from Table~\ref{tbl-2:data_of_two_TP_AGB_stars}, the massive 
quasi-steady wind from IRC+10216 can be driven by the Alfv\'en wave-driven mechanism, although $v_{\rm gas}$ 
in our model (11.4km s$^{-1}$) is somewhat lower than that derived from the observations 
(e.g., 14.6 km s$^{-1}$ from \citet{kna98}). 

The following should be noted; The slow and massive winds from C-rich AGB stars have been investigated in 
the framework of pulsation-enhanced dust-driven wind models based on the formulation by 
\citet{gai88} and \citet{gau90}. The hydrodynamical models roughly reproduce the observed dynamical behavior of 
these stars with $\dot{M}$ $\geq$ 10$^{-6}M_{\odot}$ yr$^{-1}$ \citep{win94,win97,now05}. The wind properties 
derived from our MHD simulations should be modified by taking into account the force acting on carbon grains. 
Furthermore it is a fact that, at present time, no satisfactory hydrodynamical model is available for reproducing 
the observed physical properties of dusty wind from O-rich AGB stars \citep[e.g.,][]{ohn16}. Thus the wind 
properties of dusty TP-AGB stars should be investigated by using our MHD model coupled with dust formation calculation.

\section{Summary}
The properties and the stability of the Alfv\'en wave-driven winds from the RGB and the AGB stars are investigated 
by employing the MHD model extended from the model developed by \citet{suz07}: the Joule resistivity and 
the ambipolar diffusion are included for the magnetic diffusion terms, the radiative cooling rate at low 
temperature is modified, and the radiative heating rate is 
estimated based on the radiation field derived using the Unno-Kondo method of \citet{win97}. 
The model is applied to investigate the stability and the properties of wind from the RGB and the AGB stars 
whose stellar parameters necessary for the MHD 
simulations are calculated by the MESA code for the stars with the initial mass $M_{\rm ini}$ in the range of 1.5 to 
3 $M_{\odot}$ and the initial metallicity  of $Z_{\rm ini}$ = 0.02. The results of the simulations are summarized 
as follows: 

1. The stars with $T_{{\rm eff}}$ $>$ 4530 K in the RGB and E-AGB phases exhibit 
the chromospheric structures and stable winds, regardless of the initial mass. The stable wind is 
characterized by the high gas temperature exceeding 10$^{5}$ K, the high outflow velocity of 
$v_{{\rm gas}}$ $>$ 80 km s$^{-1}$, and the low mass-loss rate of $\dot{M}$ $\leq$ 10$^{-11} M_{\odot}$ yr$^{-1}$, 
and is classified as the first type wind. The mass-loss rate is at least two orders of magnitude smaller than 
that of \citet{sch05}, which may indicate that the empirical formulae are derived based on the observations of 
the magnetically active giants with  $f_{{\rm 0}}B_{r,{\rm 0}}$ $>$ 5G.

2. In the stars with 4420 K $<$ $T_{{\rm eff}}$ $\leq$ 4530 K on the RGB and E-AGB phases, 
chromospheric structures are not always seen. The winds  
%from stars with $M_{\rm ini}$ is 1.5 and 2.0 $M_{\odot}$ are stable 
blow incessantly, but characterized by the low gas temperature lower than $T_{{\rm eff}}$, the quite 
low velocity of  $v_{{\rm gas}}$  $<$ 10 km s$^{-1}$ and the mass-loss rate  $\dot{M}$  in the range of 
5 $\times$ 10$^{-11}$--2 $\times$ 10$^{-10} M_{\odot}$ yr$^{-1}$. The wind classified as the second type 
seems not to be stable. 

3. In the regime of 3450 K $<$ $T_{{\rm eff}}$ $\leq$ 4420 K neither stable wind nor chromospheric 
structure forms in the $M_{\rm ini}$ = 1.5 $M_{\odot}$, while the atmosphere is lifted due to the 
injection of Alfv\'en wave (the third type wind). The levitation of atmosphere is typically seen in 
relatively swollen stars on the E-AGB and stars on the TP-AGB, as well at $M_{\ast}$ = 1.4267 $M_{\odot}$ 
around the tip of RGB. In these stars, other mechanisms are required to keep stable wind blowing.

4. On the TP-AGB of all model stars, the formation of stable Alfv\'en wave-driven referred to 
as the fourth type is possible because 
the damping of magnetic perturbation is more suppressed in the inner atmosphere and the gravity in the 
outer wind part gets lower with evolving the star. The stable stellar winds are slow, dense, and 
super-Alfv\'enic without any chromospheric structure\textbf{.}  

The dissipation of Alfv\'en waves injected from the stellar surface can lead to stable winds 
on the stars in the RGB and AGB phases which are plotted in HR diagram in Figure~\ref{fig9}. 
In particular, it is shown for the first time that the slow and massive wind can be magnetically 
driven without any chromospheric structure in low-gravity AGB stars. 
In the future work we will investigate systematically the dependence of the surface magnetic field 
strength on the wind properties along the stellar evolution. Furthermore we will extend our code to 
investigate how stellar pulsation and dust formation play a role in driving the wind in AGB stars.

\acknowledgments

The authors would like to thank Graham M. Harper for providing the code to calculate the 
ionization degree. This work is supported in part by Grants-in-Aid for Scientific Research from the MEXT of Japan, 
16H02168 and 17H1105.




\begin{thebibliography}{}
\bibitem[Airapetian et al.(2010)]{air10} Airapetian, V., Carpenter, K. G., Ofman, L. 2010 \apj, 723, 1210
\bibitem[Anderson \& Athay(1989)]{and89} Anderson, L. S., Athray, R. G. 1889, \apj, 336, 1089
\bibitem[Auri\`ere et al.(2009)]{aur09} Auri\`ere, M., Wade, G. A., Konstantinova-Antova, R. et al. 2009, \aap, 504, 231
\bibitem[Auri\`ere et al.(2015)]{aur15} Auri\`ere, M., Konstantinova-Antova, R., Charbonnel. C.,  et al. 2015, \aap, 574, A90
\bibitem[Belcher \& Olbert (1975)]{bel75}  Belcher, J. W., Olbert, S. 1975, \apj, 200, 369
\bibitem[Bergeat \& Chevallier(2005)]{ber05} Bergeat, J.,  Chevallier, L. 2005, \aap, 429, 235 
\bibitem[Bowen(1988)]{bow88} Bowen, G. H. 1988, \apj, 329, 299
\bibitem[Catelan(2000)]{cat00} Catelan, M.  2000, \apj, 531, 826
\bibitem[Cox \& Giuli(1968)]{cox68} Cox, J. P., Giuli, R. T. 1968, Principle of stellar structure, Gordon and Breach, NY  
\bibitem[Cristallo et al.(2009)]{cri09} Cristallo, S., Straniero, O., Gallino, R. 2009, \apj, 696, 797
\bibitem[Cristallo et al.(2011)]{cri11} Cristallo, S., Piersanti, L, Straniero, O. 2011, \apjs, 197, 17
\bibitem[Crosas \& Menten(1997)]{cro97} Cross, M., Menten, K. M. 1997, \apj, 483, 913 
\bibitem[Draine et al.(1983)]{dra83} Draine, B. T., Roberge, W. G., Dalgarno, A. 1983, \apj, 264, 485  
\bibitem[Dupree et al.(2009)]{dup09} Dupree, A. K., Smith, G. H., Strader, J., et al. 2009, \aj, 138, 1485 
\bibitem[Duthu et al.(2017)]{dut17} Duthu,  A., Herpin, F., Wiesemeyer, H., et al. 2017, \aap, 609, A12
\bibitem[Fleischer et al.(1992)]{fle92} Fleischer, A. J., Gauger, A., Sedlmayr, E. 1992, \aap, 266, 321
\bibitem[Gail \& Sedlmayr(1988)]{gai88} Gail, H.-P., \& Sedlmayr,  E. 1988, \aap, 206, 153
\bibitem[Gauger et al.(1990)]{gau90} Gauger, A., Gail, H.-P.,  Sedlmayr, E. 1990, \aap, 235, 345
\bibitem[Haisch et al.(1980)]{hai80} Haisch, B. W., Linsky, J. L., Basri, G. S. 1980, \apj, 235, 519 
\bibitem[Hartmann \& MacGregor (1980)]{har80} Hartmann, L., MacGregor, K. B. 1980, \apj, 242, 260
\bibitem[Hartmann \& Avrett(1984)]{har84} Hartmann, L., Avrett, E. H. 1984, \apj, 284, 238
\bibitem[H\"ofner \& Dorfi(1997)]{hof97}  H\"ofner, S., Dorfi, E. A. 1997 \aap, 319, 648
\bibitem[H\"ofner(2008)]{hof08}  H\"ofner, S., 2008 \aap, 491, L1
\bibitem[H\"ofner et al.(2016)]{hof16} H\"ofner, S.,Bladh, S., Aringer, B. \& Ahuja R. 2016, \aap, 594, A108 
\bibitem[Holzer et al.(1983)]{hol83} Holzer, T. E., Fla, T., Leer, E. 1983, \apj, 275, 808 
\bibitem[Ivezi\'c \& Elitzur(1996)]{ive96} Ivezi\'c,  \v{Z}., Elitzur, M. 1996, \mnras, 279, 1019
\bibitem[Judge \& Stencel(1991)]{jud91} Judge, P. G., Stencel, R. E. 1991, \apj, 371, 357
\bibitem[Knapp et al.(1998)]{kna98} Knapp, G. R., Young, K,, Lee E., Jorissen A. 1998, \apjs, 117, 209 
\bibitem[Kopp \& Holzer(1976)]{kop76} Kopp, R.A., Holzer, T.E., 1976, Sol. Phys., 39, 43
\bibitem[Kozasa \& Sogawa(1997)]{koz97} Kozasa, T., Sogawa, H. 1997, \apss, 251, 165 
\bibitem[Kozasa \& Sogawa(1998)]{koz98} Kozasa, T., Sogawa, H. 1998, \apss, 255, 437 
\bibitem[Ladjal et al.(2010)]{lad10} Ladjal, D., Barlow, M. J., Groenewegen, M. A. T., et al. 2010, \aap, 518, L141 
\bibitem[L\`ebre et al.(2014)]{leb14} L\`ebre, A., Auri\`ere, M., Fabas, N., et al. 2014, \aap, 561, A85
\bibitem[Linsky \& Haisch(1979)]{lin79} Linsky, J. L., Haisch, B. M. 1979 \apj, 229, L27 
\bibitem[Marigo(2002)]{mar02} Marigo, P., 2002. \aap, 387, 507
\bibitem[Marigo \& Aringer(2009)]{mar09} Marigo, P., Aringer, B. 2009, \aap, 508, 1539
\bibitem[Marigo et al.(2013)]{mar13} Marigo, P., Bressan, A., Mammi, A. 2013, \mnras, 434, 488   
\bibitem[Mathis et al.(1983)]{mat83} Mathis, J. S., Mezger, P. G., Panagia, N. 1983, \aap, 128, 212
\bibitem[Menten et al.(2012)]{men12} Menten, K. M., Reid, M J., Kami\'nski, T., Claussen, M. J. 2012 \aap, 543, A73
\bibitem[M\'esz\'aros et al.(2009)]{mes09} M\'esz\'aros, S., Avrett, E. H., Dupree, A. K. 2009, \aj, 138, 615
\bibitem[McDonald \& Zijlstra(2015)]{mcd15} McDonald, I., Zijlstra, A. A.,2015, \mnras, 448, 502
\bibitem[Moriyasu et al.(2004)]{mor04} Moriyasu, S., Kudoh, T., Yokoyaman, T, Shibata, K. 2004, \apjl, 601, L107
\bibitem[Nowotny et al.(2005)]{now05} Nowotny, W., Lebzelter, T., Hron, J., \& H\"ofner, S. 2005, \aap, 437, 285 
\bibitem[O'Gorman et al.(2017)]{ogo17} O'Gorman, E., Harper, G. M., Vlemmings, W. 2017, \aap, 599, A47
\bibitem[Ohnaka et al.(2016)]{ohn16} Ohnaka, K., Weigelt, G., Hofmann, K.-H. 2016, \aap, 589, A91
\bibitem[Paxton et al.(2011)]{pax11} Paxton, B., Bildsten, L.,  Dotter, A., et al. 2011, \apjs, 192, 3 
\bibitem[Paxton et al.(2013)]{pax13} Paxton, B., Cantiello, M.,  Arras, P., et al. 2013, \apjs, 208, 4 
\bibitem[Paxton et al.(2015)]{pax15} Paxton, B., Marchant, P.,  Schwab, J., et al. 2015, \apjs, 220, 15 
\bibitem[Perrin et al.(2004)]{per04} Perrin, G., Ridgeway, S. T., Mennesson, B., et al. 2004, \aap, 426, 279 
\bibitem[Ramstedt \& Olofsson(2014)]{ram14} Ramstedt, S., Olofsson, H. 2014, \aap, 566, A145
\bibitem[Reimers(1975)]{rei75} Reimers, D., 1975, Mem. Toy. Soc. Li\`ege  8, 369 
\bibitem[Renzini et al.(1977)]{ren77} Renzini, A., Cacciari, C., Ulmschneider, P., Schmitz, F. 1977 \aap, 61, 39 
\bibitem[Rosenfield et al.(2014)]{ros14} Rosenfield, P.,  Marigo, P.,  Girardi, L., et al. 2014, \apj, 790, 22
\bibitem[Rosenfield et al.(2016)]{ros16} Rosenfield, P.,  Marigo, P.,  Girardi, L., et al. 2016, \apj, 822, 73
\bibitem[Sandin(2008)]{san08} Sandin, C.  2008, \mnras, 385, 215
\bibitem[Schirrmacher et al.(2003)]{sch03} Schirrmacher, V.,  Woitke, P., \& Sedlmayr, E. 2003, \aap, 404, 267
\bibitem[Sch\"oier et al.(2013)]{sch13} Sch\"oier, F. L., Ramstedt, S., Olofsson, H., et al, 2013, \aap, 550, A78
\bibitem[Schr\"oder \& Cuntz(2005)]{sch05} Schr\"oder, K.-P.,  Cuntz, M. 2005, \apjl, 630, L73
\bibitem[Sudou et al.(2017)]{sud17} Sudou, H., Shiga, M., Omodaka, T., et al. 2017, JKAS , 50, 157
\bibitem[Sutherland \& Dopita(1993)]{sut93} Sutherland, R. S., Dopita, M. A. 1993, \apjs, 88, 253
\bibitem[Suzuki(2007)]{suz07} Suzuki, T. K. 2007, \apj, 659, 1592
\bibitem[Suzuki \& Inutsuka (2005)]{suz05} Suzuki, T. K., Inutsuka, S. 2005, \apjl, 632, L49
\bibitem[Suzuki \& Inutsuka (2006)]{suz06} Suzuki, T. K., Inutsuka, S. 2006, J. Geophys. Res., 111, A06101 
\bibitem[Suzuki et al.(2013)]{suz13} Suzuki, T. K., Imada, S., Kataoka, R. et al. 2013, \pasj, 65, 98 
\bibitem[Tashibu et al.(2017)]{tas17} Tashibu, S., Yasuda, Y.,  Kozasa, T. 2017,  \mnras, 466, 1709
\bibitem[Thirumalai \& Heyl(2013)]{thi13} Thirumalai, A., Heyl, J. S 2013, \mnras, 430, 1359
\bibitem[Ventura \& Marigo(2010)]{ven10} Ventura, P., Marigo, P. 2010, \mnras, 408, 2476 
\bibitem[Ventura et al.(2012)]{ven12} Ventura, P., Di Criscienzo, M., Schneider. R., 2012, \mnras, 420, 1442 
\bibitem[Ventura et al.(2018)]{ven18} Ventura, P., Karakas, A., Dell'Agli, F., et al. 2018, \mnras, 475, 2282
\bibitem[Vidotto \& Jatenco-Pereira(2006)]{vid06} Vidotto, A. A.,  Jatenco-Pereira, V. 2006, \apj, 639, 416 
\bibitem[Weiss \& Ferguson(2009)]{wei09} Weiss, A., Ferguson, J. W, 2009, \aap, 508, 1343
\bibitem[Winters et al.(1994)]{win94} Winters, J. M., Dominik, C., Sedlmayr, E. 1994, \aap, 288, 255
\bibitem[Winters et al.(1997)]{win97} Winters, J. M., Fleischer, A. J., Le Bertre, T.,  Sedlmayr, E. 1997, \aap, 326, 305
\bibitem[Woitke et al.(1996)]{woi96} Woitke, P., Kr\"uger, D., Sedlmayr, E. 1996, \aap, 311, 927 
\bibitem[Wood(1979)]{woo79} Wood, P. R. 1979, \apj, 227, 220
\bibitem[Wood et al.(2016)]{woo16} Wood, B. E., M\"uller,  H.-R., Harper, G. M. 2016,  \apj, 829, 74 
\bibitem[Woodruff et al.(2004)]{woo04} Woodruff, H. C., Eberhardt, M., Driebe, T., et al. 2004, \aap, 421, 703 
\bibitem[Wyatt \& Cahn(1983)]{wya83} Wyatt,  S. P., Cahn, J. H., 1983 \apj, 275, 225  
\bibitem[Yasuda \& Kozasa(2012)]{yas12} Yasuda, Y.,   Kozasa, T. 2012, \apj, 745, 159
\bibitem[Young(1995)]{you95} Young, K. 1995, \apj, 445, 872
\end{thebibliography}
\end{document}